\documentclass[12pt]{article}

\baselineskip=\normalbaselineskip

\usepackage{mathrsfs,amsbsy,amssymb,latexsym,amsfonts,amsmath}
\usepackage{graphicx,color}

\allowdisplaybreaks[2]
\renewcommand{\thefootnote}{\arabic{footnote}}
\makeatletter
\catcode`\@=11
\@addtoreset{equation}{section}

\def\@seccntformat#1{\csname the#1\endcsname.~~}
\makeatother
\newcommand{\nn}{\nonumber}
\newcommand{\rmd}{{\mathrm d}}
\newcommand{\tr}{\mathrm{Tr}}
\newcommand{\gyms}{g^2_{\mathrm{YM}}}
\newcommand{\da}{D^{(a)}}
\newcommand{\sa}{\sigma_0^{\mathrm{adj}}}
\newcommand{\tm}{\tilde{\mu}}
\begin{document}
%%%%%%%%%%%%%%%%%%%%%%%%%%%%%%%%%%%%%%%%%%%%%%%%%%%%%%%%%%%%%

%\baselineskip 0.7cm

\begin{titlepage}

%% Set the number of the title with 0
\setcounter{page}{0}

%% change the footnote symbol
\renewcommand{\thefootnote}{\fnsymbol{footnote}}

\begin{flushright}
%{\tt 
KUNS-2458 \\
YITP-13-72 \\
%\\}
\end{flushright}

\vskip 1.35cm

\begin{center}
{\Large \bf 
Exact Results in Supersymmetric Field Theories 
on Manifolds with Boundaries
}

\vskip 1.2cm 

{\normalsize
Sotaro Sugishita${}^a$\footnote{sotaro(at)gauge.scphys.kyoto-u.ac.jp} 
and Seiji Terashima${}^b$\footnote{terasima(at)yukawa.kyoto-u.ac.jp}
}

\vskip 0.8cm
${}^a${\it Department of Physics, Kyoto University, Kyoto 606-8502, Japan}

${}^b${\it Yukawa Institute for Theoretical Physics, Kyoto University, Kyoto 606-8502, Japan
}

\end{center}

\vspace{12mm}

\centerline{{\bf Abstract}}

We construct supersymmetric gauge theories on some curved manifolds 
with boundaries. 
Our examples include a part of three-sphere and
a part of two-sphere.
We concentrate on 
Dirichlet boundary conditions.
For these theories on the manifolds with the boundaries,
we compute the partition functions and the Wilson loops 
exactly using the localization technique.

\end{titlepage}
\newpage

\tableofcontents
\vskip 1.2cm

\section{Introduction and Summary}

In order to understand 
the dynamics of the gauge theory,
we need to perform non-perturbative computations
reliably.
The analytical or exact results of them
are especially important although 
they may be extraordinary difficult
for generic gauge theories.
The gauge theories with the supersymmetry (SUSY)
are highly non-trivial examples 
of the theories in which exact results are obtained.
The SUSY gauge theories themselves are important objects
related to string theory, mathematics and many other areas.
Even for the understanding the non-SUSY gauge theories,
the SUSY gauge theories will be important because 
of the exact results and the physical consequences from them.

The exact results for the SUSY gauge theories
have been mainly obtained using the holomorphy \cite{Seiberg,SW,IS},
which imposes strong constraints for the certain quantities.
Recently, the more direct computations of the partition function
and other operators using the so called localization technique
has been extensively developed \cite{Nekrasov, Pestun}.
In particular, in \cite{Pestun}, the 4D ${\cal N}=2$ SUSY gauge theories 
on a curved space were constructed and then the partition function and 
the Wilson loops are exactly computed using the localization technique. 
This method has advantages of the wide applicability and 
the straightforwardness.
It has been applied for the SUSY gauge 
theories on various curved spaces with different dimensions 
\cite{Kim2}-\cite{Kim:2013nva}, however,
the SUSY gauge theories on manifolds with boundaries 
have never been investigated using the localization technique.
The exact results in such theories will be also 
highly important, for example, if we remember the 
recent application of the SUSY theories on $S^2$ as 
string world sheet actions \cite{Romo}.

In this paper,
we explicitly construct the 
supersymmetric gauge theories on some curved manifolds with
boundaries.\footnote{
The SUSY theories with boundaries on flat spaces have been
considered in, for examples, \cite{HIV, Hori, HHP} for 2D, 
\cite{Berman, Berman:2009xd, Yamaguchi, Vassilevich:2003xk} for 3D and \cite{GW}
for 4D.}
Our examples include a part of $S^3$ with a torus boundary
and a part of $S^2$ with a $S^1$ boundary. 
We use off-shell formulation of the SUSY and
concentrate on 
Dirichlet boundary conditions.
Actually, 
there seems to be no consistent Neumann-like boundary conditions
with the SUSY if we impose the boundary condition 
which eliminates only the half of the fermions.\footnote{
This assumption may be too strong. Indeed, it may be 
possible to use 
the boundary condition used in \cite{Hori}
even in the off-shell formulation 
or to introduce some boundary degrees of freedom, 
however, we leave such possibilities for future works.
}
For the SUSY theories,
we compute the partition functions and the Wilson loops 
exactly using the localization technique.

Needless to say, our work is just a first step
and there are many points which need further investigations:
for examples, 
the interpretations and the applications of our results,
the other boundary conditions,
and extensions to other topologies and other dimensions.
One important direction is to extend
the ABJM model with the boundary condition 
representing the M5-branes \cite{Te2, GRVV} 
to the curved manifold and compute some exact quantities
because these might give us a clue for the M5-branes.
We hope to report for the investigations of 
these topics in the future.

The organization of this paper is as follows:
In section \ref{3D-th}, 
we construct 3D SUSY field theories 
on the manifold with the torus boundary, which
is constructed by cutting an $S^3$.
The consistent boundary conditions we impose 
are the Dirichlet boundary conditions.
Using the localization technique,
we compute the partition function and the Wilson loop 
of the theories exactly.
In section \ref{2D-th}, 
we construct 2D SUSY field theories 
on the manifold with the circle boundary, which
is constructed by cutting an $S^2$.
The partition function and the Wilson loop 
for the theories are computed exactly.

The notations and some useful formulas are summarized in the Appendix \ref{notation}.

Note added: 
The authors thank K. Hori and M. Romo 
and D. Honda and T. Okuda
for notifying them of submissions of related papers\cite{Honda:2013uca, Hori:2013ika}.

%%%%%%%%%%%%%%%%%%%%%%%%%%%%%%%%%%%%%%%%%%%%%%%%%%%%%%%%%%%%%

\section{Three-dimensional theories}
\label{3D-th}

\subsection{A 3D manifold with a boundary}

We will describe
a three dimensional manifold with a boundary
on which the supersymmetric field theories constructed. 
First we recall the round $S^3$, i.e. 
(${X_0}^2+{X_1}^2+{X_2}^2+{X_3}^2=\ell^2$ in $\mathbb{R}^4$).
The coordinates we will use are
\begin{align}
\begin{split}
X_0&=\ell\cos\theta\cos\varphi\,,
X_1=\ell\cos\theta\sin\varphi\,,\\
X_2&=\ell\sin\theta\cos\chi\,,
X_3=\ell\sin\theta\sin\chi\,.
\end{split}
\end{align}
where
$0\leq\theta\leq\pi/2\,,0\leq\varphi\leq 2\pi\,,0\leq\chi\leq 2\pi$.
The metric is given by
\begin{align}
\rmd s^2 &= \ell^2(\rmd \theta^2 +\cos^2\theta \rmd \varphi^2 
+\sin^2\theta \rmd \chi^2)\,,\\
\sqrt{g}&=\ell^3 \cos\theta \sin\theta\,.
\end{align}
We take the following dreibein:
\begin{align}
e^1=\ell\cos\theta\,\rmd\varphi\,,\quad
e^2=\ell\sin\theta\,\rmd\chi\,,\quad
e^3=\ell\,\rmd\theta\,.
\end{align}
and the gamma matrices:
\begin{align}
\gamma^\varphi =\frac{1}{\ell\cos\theta}\gamma^1
\,,\quad
\gamma^\chi =\frac{1}{\ell\sin\theta}\gamma^2
\,,\quad
\gamma^\theta =\frac{1}{\ell}\gamma^3 \,.
\end{align}

Then, the manifold with the boundary is defined by
just restricting the coordinate $\theta$ as
\begin{eqnarray}
 0\leq\theta\leq\theta_0\,,
\end{eqnarray}
where
$0<\theta_0\leq\pi/2$.
Thus the boundary defined by $\theta=\theta_0$ is 
a torus parameterized by $\phi$ and $\chi$, except for $\theta_0=\pi/2$.
We will see that
on the manifold with the boundary we can construct
field theories with two supersymmetries (which are the half 
or $1/4$ of 
the supersymmetries of the ${\cal N}=2$ supersymmetric filed theories
on the round $S^3$).

%We cut the $S^3$ by $\theta=\theta_0$ 
%and consider the region $0\leq\theta\leq\theta_0$\,.\\
%boundary:\quad $\theta=\theta_0\quad(0<\theta_0\leq\pi/2)$

\subsection{3D supersymmetric field theories}

Now we will construct the supersymmetric field theories
on the manifold with the boundary.

First, we will summarize the SUSY transformations and 
the SUSY invariant Lagrangian of 
the ${\cal N}=2$ supersymmetric field theories
on the round $S^3$ \cite{Kapustin:2009kz, Jafferis, Hama-HL}.
We will write the derivative terms in the Lagrangians such that
they are SUSY invariant on the manifold with the boundary. 
The Killing spinors on $S^3$ ware given by\footnote{
There are the positive and the negative Killing spinors:
$D_{\mu}\epsilon = \pm\frac{i}{2\ell}\gamma_{\mu}\epsilon$.
Either ones lead same result, thus we take the positive ones.}
\begin{align}
D_{\mu}\epsilon = \frac{i}{2\ell}\gamma_{\mu}\epsilon,
\end{align}
which is solved as:
%$( D_{\mu}\epsilon =i/(2\ell)\gamma_{\mu}\epsilon) $ :
\begin{align}
\epsilon=\frac{C_1}{\sqrt{2}}
\begin{pmatrix}
-e^{-\frac{i}{2}(\varphi-\chi-\theta)}\\
e^{-\frac{i}{2}(\varphi-\chi+\theta)}
\end{pmatrix} 
+\frac{C_2}{\sqrt{2}}
\begin{pmatrix}
e^{\frac{i}{2}(\varphi-\chi+\theta)}\\
e^{\frac{i}{2}(\varphi-\chi-\theta)}
\end{pmatrix}
\,, 
\end{align} 
in our basis.

The SUSY transformations of the vector multiplets 
with the Grassmann odd Killing spinor parameters $\epsilon, \bar{\epsilon}$
are the followings:
\begin{align}
\begin{split}
\label{susytr1}
\delta A_{\mu} 
&=-\frac{i}{2}(\bar{\epsilon}\gamma_{\mu}\lambda-\bar{\lambda}\gamma_{\mu}\epsilon)\,, \quad
\delta\sigma 
=\frac{1}{2}(\bar{\epsilon}\lambda-\bar{\lambda}\epsilon)\,,
\\
\delta\lambda
&=\Bigl(
\frac{1}{2}\gamma^{\mu\nu}F_{\mu\nu}-D+i D_{\mu}\sigma \gamma^{\mu}
\Bigr)\epsilon
+\frac{2i}{3}\sigma\gamma^{\mu}D_{\mu}\epsilon\,,
\\
\delta\bar{\lambda} 
&=\Bigl(
\frac{1}{2}\gamma^{\mu\nu}F_{\mu\nu}+D-i D_{\mu}\sigma \gamma^{\mu}
\Bigr)\bar{\epsilon}
-\frac{2i}{3}\sigma\gamma^{\mu}D_{\mu}\bar{\epsilon}\,,
\\
\delta D 
&= -\frac{i}{2}\bar{\epsilon}(\gamma^{\mu}D_{\mu}\lambda-[\lambda,\sigma])
-\frac{i}{2}(D_{\mu}\bar{\lambda}\gamma^{\mu}-[\bar{\lambda},\sigma])\epsilon
-\frac{i}{6}(D_{\mu}\bar{\epsilon}\gamma^{\mu}\lambda
+\bar{\lambda}\gamma^{\mu}D_{\mu}\epsilon)\,,
\end{split}
\end{align}
where all the fields are in the adjoint representation of the 
gauge group $G$ although we have not written the indices for it explicitly.
Note that 
$a$ and $\bar{a}$ are independent fields for the Grassmann odd fields
although usually $\bar{a}$ means a (complex) conjugate.

For the chiral multiplet of R-charge $q$, 
the SUSY transformations are:
\begin{align}
\begin{split}
\delta\phi &= \bar{\epsilon} \psi \,,\qquad
\delta \psi 
= \Bigl( i D_{\mu} \phi\gamma^{\mu} + i \sigma\phi \Bigr) \epsilon 
+\frac{2i q}{3}\phi\gamma^{\mu}D_{\mu}\epsilon+ \bar{\epsilon} F \,,\\
\delta \bar{\phi} &= \epsilon \bar{\psi} \,,\qquad
\delta \bar{\psi} 
= \Bigl( i D_{\mu}\bar{\phi}\gamma^{\mu} +i \bar\phi\sigma \Bigr) \bar{\epsilon} 
+\frac{2i q}{3}\bar{\phi}\gamma^{\mu}D_{\mu}\bar{\epsilon}+ \epsilon \bar{F} \,,\\
\delta F &=
\epsilon \Bigl( i \gamma^{\mu} D_{\mu} \psi - i \sigma \psi -i\lambda \phi \Bigr) 
+\frac{i}{3}(2q-1)D_{\mu}\epsilon \gamma^{\mu}\psi\,,\\
\delta \bar{F} &=
\bar{\epsilon} \Bigl( i \gamma^{\mu}D_{\mu}\bar{\psi} -i \bar{\psi}\sigma
+i\bar{\phi} \bar{\lambda} \Bigr)
+\frac{i}{3}(2q-1)D_{\mu}\bar{\epsilon} \gamma^{\mu}\bar{\psi} \;,
\label{susytr2}
\end{split}
\end{align}
where $\phi, \psi, F$ are in a representation $R$ of $G$
and $\bar{\phi}, \bar{\psi}, \bar{F}$ are in 
the complex conjugate representation of $R$.

There are several invariant actions under the 
SUSY transformations \eqref{susytr1}-\eqref{susytr2}.
The one is the Yang-Mills Lagrangian:\footnote{Note that 
the fermion kinetic terms are symmetrical with respect to 
$\lambda$ and $\bar\lambda$. In the other Lagrangians, we
also use the symmetric fermion kinetic terms.}
\begin{align}
\frac{1}{\gyms}\mathcal{L}_{\mathrm{YM}}
&=\frac{1}{\gyms}\tr\Bigl(
\frac{1}{4}F_{\mu\nu}F^{\mu\nu}+\frac{1}{2}D_{\mu}\sigma D^{\mu}\sigma
+\frac{1}{2}(D+\sigma /\ell)^2\nn\\
&\qquad\qquad\quad +\frac{i}{4}\bar{\lambda}\gamma^{\mu}D_{\mu}\lambda
+\frac{i}{4}\lambda\gamma^{\mu}D_{\mu}\bar{\lambda}
+\frac{i}{2}\bar{\lambda}[\sigma,\lambda]-\frac{1}{4\ell}\bar{\lambda}\lambda
\Bigr)\,,
\end{align}
where $g_{\mathrm{YM}}$ is the coupling constant.
Another one is the Chern-Simons term:
\begin{align}
\mathcal{L}_{\mathrm{CS}}&=
i \frac{k}{4\pi}\tr\Bigl(\varepsilon^{\mu\nu\lambda}
(A_\mu\partial_\nu A_\lambda-\frac{2 i}{3}A_\mu A_\nu A_\lambda)
-\bar{\lambda}\lambda+ 2 D\sigma\Bigr)\,,
\end{align}
where the level $k$ is an integer for the theory on $S^3$.
We can also construct the Fayet Iliopoulos (FI) term:
\begin{align}
\mathcal{L}_{\mathrm{FI}}&=
\frac{i\zeta}{\pi\ell}\tr(D-\sigma /\ell)\,.
\end{align}
For the chiral multiplets, the matter kinetic terms
are given by
%\footnote{Note that 
%the fermion kinetic terms are symmetrical with respect to 
%$\psi$ and $\bar\psi$.}
\begin{align}
\mathcal{L}_{\mathrm{mat}}
&=D_{\mu}\bar{\phi}D^{\mu}\phi+\bar{\phi}\sigma^2\phi
+\frac{i(2q-1)}{\ell}\bar{\phi}\sigma\phi+\frac{q(2-q)}{\ell^2}\bar{\phi}\phi
+i\bar{\phi}D\phi+\bar{F}F\nn\\
&\quad-\frac{i}{2}\bar{\psi}\gamma^{\mu}D_{\mu}\psi
+\frac{i}{2}D_{\mu}\bar{\psi}\gamma^{\mu}\psi
+i\bar{\psi}\sigma\psi-\frac{(2q-1)}{2\ell}\bar{\psi}{\psi}
+i\bar{\psi}\lambda\phi -i\bar{\phi}\bar{\lambda}\psi\,.
\end{align}
We can also have the superpotential terms (without derivatives)
which are same form as the ones in the flat space.
Note that the bosonic kinetic terms for the chiral multiplet
are written as
\begin{align}
&D_{\mu}\bar{\phi}D^{\mu}\phi+\bar{\phi}\sigma^2\phi
+\frac{i(2q-1)}{\ell}\bar{\phi}\sigma\phi+\frac{q(2-q)}{\ell^2}\bar{\phi}\phi
+i\bar{\phi}D\phi+\bar{F}F
 +\frac{1}{2 g_{YM}^2}(D+\sigma /\ell)^2 \nn  \\
&= \left| D_\mu \phi \right|^2 
+\bar{\phi} (\sigma+\frac{i}{\ell}(q-1))^2 \phi
+\frac{1}{\ell^2} |\phi|^2 
+\frac{1}{2} \left(
\frac{1}{g_{YM}} (D+\sigma /\ell)+
i g_{YM} \phi \bar{\phi} \right)^2
+\frac{g_{YM}^2}{2} |\phi|^4
+|F|^2,
\end{align}
which can be positive definite
if we shift $\sigma, D$ appropriately.

\subsubsection{The boundary condition}

If one imposes the boundary conditions on fields,
the supersymmetry which preserves these boundary conditions is remain.
There are possible candidates for 
consistent supersymmetric boundary conditions for our theories.
It is an important question which ones indeed works.
In this paper, however, we study the Dirichlet 
condition only as a first step.

Denoting $\tm$ as the coordinates tangent to the boundary, 
i.e. $\{ \varphi, \chi \}$,
our ansatz for the boundary conditions for the vector multiplet is:
\begin{align}
A_{\tm} |_{\theta=\theta_0} = a_{\tm} \, ,
%\text{const.}&=\text{Cartan}
%\equiv 
\label{3d-bc-a}\\
\sigma |_{\theta=\theta_0} = %\text{const.}&=\text{Cartan}
%\equiv 
\sigma_0\,,
\label{3d-bc-s}\\
\ell e^{i(\varphi-\chi)}\gamma^\theta\lambda |_{\theta=\theta_0} 
&=\bar\lambda |_{\theta=\theta_0} \,,
\label{3d-bc-la}
\end{align}
where $a_{\tilde{\mu}}$ and $\sigma_0$ are constants and in the Cartan part of the
adjoint representation. 
We do not impose any conditions for other fields ( $A_\theta$ and $D$ ).
Note that the conditions for the bosons may 
be covariantly represented as 
$F_{\tm \tilde{\nu}}|_{\theta=\theta_0}=0$ and $D_{\tm} \sigma|_{\theta=\theta_0}=0$.
We can see that 
these boundary conditions indeed
preserve the half of the supersymmetries which are generated by
the Killing spinors satisfying the relation
\begin{align}
-\ell e^{i(\varphi-\chi)}\gamma^\theta\epsilon =\bar\epsilon.
\label{3d_e_be}
\end{align}
In appendix \ref{susy-vari},
we show that the actions for the vector multiplet
are invariant under these supersymmetries with the above boundary conditions.

For chiral multiplets, we take the following boundary condition:
\begin{align}
\begin{split}
\label{3d_chiral_bc}
\phi |_{\theta=\theta_0}&=0\,,\quad
e^{\frac{i\theta_0}{2}\gamma^3}\gamma^1 e^{-\frac{i\theta_0}{2}\gamma^3}
\psi |_{\theta=\theta_0}=\psi |_{\theta=\theta_0} \,,
\\
\bar\phi |_{\theta=\theta_0}&=0\,,\quad
e^{\frac{i\theta_0}{2}\gamma^3}\gamma^1 e^{-\frac{i\theta_0}{2}\gamma^3}
\bar\psi |_{\theta=\theta_0}=-\bar\psi |_{\theta=\theta_0} \,. 
\end{split}
\end{align}
Then these boundary conditions for the chiral multiplet
preserve another half of the supersymmetries which is generated by
the Killing spinors satisfying the relations
\begin{align}
e^{\frac{i\theta}{2}\gamma^3}\gamma^1 e^{-\frac{i\theta}{2}\gamma^3}
\epsilon =-\epsilon \,,\quad
e^{\frac{i\theta}{2}\gamma^3}\gamma^1 e^{-\frac{i\theta}{2}\gamma^3}
\bar\epsilon =\bar\epsilon \,.
\label{3d_e_be 2}
\end{align}
We also show that the matter kinetic action is invariant under the SUSY
with the boundary condition in Appendix \ref{susy-vari}.

We will consider SUSY theories with both vector and chiral multiplets,
thus only the $1/4$ SUSY will remain.
The Grassmann even Killing spinors 
satisfying the both relations are given by
\begin{align}
\epsilon = \frac{1}{\sqrt{2}}
\begin{pmatrix}
-e^{-\frac{i}{2}(\varphi-\chi-\theta)}\\
e^{-\frac{i}{2}(\varphi-\chi+\theta)}
\end{pmatrix} 
\quad 
\text{and} 
\quad
\bar\epsilon =
\frac{1}{\sqrt{2}}
\begin{pmatrix}
e^{\frac{i}{2}(\varphi-\chi+\theta)}\\
e^{\frac{i}{2}(\varphi-\chi-\theta)}
\end{pmatrix}
\, .
\label{bks3}
\end{align}
We can compute the bi-linears of the Grassmann even spinors:
\begin{align}
\bar\epsilon \epsilon &=1 \,,
\\
\bar\epsilon\gamma^a \epsilon &= (-\cos\theta, \sin\theta, 0)
\equiv v^a\,,
\label{v}
\\
\epsilon\gamma^a \epsilon &= (i\sin\theta, i\cos\theta, 1)e^{-i(\varphi-\chi)}
\equiv v^a_+\,,
\label{vplus}
\\
\bar\epsilon\gamma^a \bar\epsilon &= (i\sin\theta, i\cos\theta, -1)e^{+i(\varphi-\chi)}
\equiv v^a_-\,,
\label{vminus}
\end{align}
which will be used later.

We should check also that 
the boundary conditions are consistent with
the variational principle.
The surface terms from variation of the Yang-Mills action are
\begin{align}
\frac{1}{\gyms}\int_{\theta=\theta_0}\rmd\varphi\,\rmd\chi\Bigl[\sqrt{g}\,\tr\Bigl(
\delta A_\nu F^{\theta\nu}
+\delta\sigma D^{\theta}\sigma
+\frac{i}{4}\bar{\lambda}\gamma^{\theta}\delta\lambda
+\frac{i}{4}\lambda\gamma^{\theta}\delta\bar{\lambda}\Bigr)\Bigr] \,,
\end{align}
which indeed vanish for the boundary conditions.
Note that we do not need to introduce boundary terms
because of our choice of the kinetic terms of the Lagrangian.
We can see that the surface terms from variation of 
the Chern-Simons action,
\begin{align}
i \int_{\theta=\theta_0}\rmd\varphi\,\rmd\chi\Bigl[
\frac{k}{4\pi}\tr(A_\chi\delta A_\varphi-A_\varphi\delta A_\chi)\Bigr]\,,
\end{align}
and the matter kinetic terms,
\begin{align}
\int_{\theta=\theta_0}\rmd\varphi\,\rmd\chi
\Bigl[\sqrt{g}\Bigl(\delta\bar\phi\,D^\theta\phi+D^\theta\bar\phi\,\delta\phi
-\frac{i}{2}\bar\psi\gamma^\theta\delta\psi
+\frac{i}{2}\delta\bar\psi\gamma^\theta\psi\Bigr)\Bigr] \,,
\end{align}
vanish for the boundary conditions.

%%%%%%%%%%%%%%%%%%%%%%%%%%%%%%%%%%%%%%%%%%%%%%%%%%%%%%%%%%%%%

%\subsection{$\delta$-exact terms}
\subsection{Localization}

In order to compute the partition function exactly
using the localization technique,
we will introduce the $\delta$-exact Lagrangian $\delta V$ as 
in \cite{Pestun}.
Note that we need to keep the total divergence terms
which have been neglected for the theories 
on the manifold without boundaries.
The localization technique implies that
the expectation values of the $\delta$-closed operators are not
changed by deforming the action
$S\to S+ t\int \delta V$,
and in the $t\to\infty$ limit,
the saddle point approximation becomes exact.

\subsubsection{3D vector multiplet}

For the vector multiplet,
we take the following $\delta$-exact term (ignoring the trace of the gauge 
indices for notational convenience):
\begin{align}
 \delta V_{\mathrm{vector}}=\frac{1}{4}\,\delta((\delta'\lambda)^{\dagger}\lambda
 + \bar\lambda(\delta'\bar\lambda)^{\dagger})\,,
\label{dV}
\end{align}
where 
$\delta, \delta'$ are the same SUSY transformations
with the Grassmann odd Killing spinors $\epsilon,\bar\epsilon$.
More precisely, we will add $t \int \delta V_{\mathrm{vector}}$
with
the Grassmann even Killing spinors
to the action.
For the computational convenience, we use the Grassmann odd spinors.
After the computations, we will replace, for examples, 
${\bar{\epsilon}}' \epsilon  = -\bar{\epsilon} \epsilon' \rightarrow 1$ 
and
${\bar{\epsilon}}' \gamma^\mu \epsilon  
= - \bar{\epsilon} \gamma^\mu \epsilon'
\rightarrow v^\mu$.
We have also defined 
\begin{align}
 (\delta\lambda)^{\dagger}&=\bar\epsilon\Bigl(
 -\frac{1}{2}\gamma^{\mu\nu}F_{\mu\nu}-D-i \gamma^{\mu}D_{\mu}\sigma-\frac{\sigma}{\ell}
 \Bigr)\,,\\
 (\delta\bar\lambda)^{\dagger}&=\Bigl(
 -\frac{1}{2}\gamma^{\mu\nu}F_{\mu\nu}-D+i D_{\mu}\sigma\gamma^{\mu}-\frac{\sigma}{\ell}
 \Bigr)\epsilon\,.
% (\delta\bar\lambda)^{\dagger}&=\Bigl(
% \frac{1}{2}\gamma^{\mu\nu}F_{\mu\nu}+D-i
% D_{\mu}\sigma\gamma^{\mu}+\frac{\sigma}{\ell}
% \Bigr)\epsilon\,.
\end{align}
With these definitions, we find that
$(\delta\lambda)^{\dagger}\delta\lambda = 
\sum_\alpha |(\delta \lambda)_\alpha|^2$,
which is manifestly positive definite.
For the bosonic part of the $\delta$-exact term, 
we can show that
\begin{align}
 (\delta'\lambda)^{\dagger}\delta\lambda 
&=\bar\epsilon'\epsilon\Bigl(
  \frac{1}{2}F_{\mu\nu}F^{\mu\nu}+D_{\mu}\sigma D^{\mu}\sigma+(D+\sigma/\ell)^2
  +\varepsilon^{\mu\nu\rho}F_{\mu\nu}D_{\rho}\sigma
 \Bigr)\,,\\
 \delta\bar\lambda(\delta'\bar\lambda)^{\dagger}  &
%=\bar\epsilon\epsilon'\Bigl(
%  \frac{1}{2}F_{\mu\nu}F^{\mu\nu}+D_{\mu}\sigma D^{\mu}\sigma+(D+\sigma/\ell)^2
%  -\varepsilon^{\mu\nu\rho}F_{\mu\nu}D_{\rho}\sigma
% \Bigr)\nn\\
 = \bar\epsilon'\epsilon\Bigl(
  \frac{1}{2}F_{\mu\nu}F^{\mu\nu}+D_{\mu}\sigma D^{\mu}\sigma+(D+\sigma/\ell)^2
  -\varepsilon^{\mu\nu\rho}F_{\mu\nu}D_{\rho}\sigma
 \Bigr)\,,
\end{align}
where we have used $\bar\epsilon\epsilon'=-\bar\epsilon'\epsilon$. 
%by \eqref{3d_e_be}.
Therefore,
\begin{align}
  (\delta'\lambda)^{\dagger}\delta\lambda+\delta\bar\lambda(\delta'\bar\lambda)^{\dagger} 
  &=4\bar\epsilon'\epsilon\Bigl(
  \frac{1}{4}F_{\mu\nu}F^{\mu\nu}+\frac{1}{2}D_{\mu}\sigma D^{\mu}\sigma
  +\frac{1}{2}(D+\sigma/\ell)^2
  \Bigr)\\
  &=4\bar\epsilon'\epsilon \,\mathcal{L}_{\mathrm{YM}}^{\mathrm{boson}}\,.
\end{align}
For the fermionic part of the $\delta$-exact term, we can show that
\begin{align}
 \delta((\delta'\lambda)^{\dagger})\lambda 
 =-\frac{1}{2}\delta F_{\mu\nu}\,\bar\epsilon'\gamma^{\mu\nu}\lambda
 -i\delta(D_\mu\sigma)\bar\epsilon'\gamma^{\mu}\lambda
 -(\delta D+\delta\sigma/\ell)\bar\epsilon'\lambda\,,
\end{align}
where
\begin{align}
 \delta F_{\mu\nu}&=\delta(\partial_\mu A_\nu-\partial_\nu A_\mu-i[A_\mu,A_\nu])\nn\\
 &=-\frac{i}{2}(\bar\epsilon\gamma_\nu D_\mu\lambda-\bar\epsilon\gamma_\mu D_\nu\lambda
 +\epsilon\gamma_\nu D_\mu\bar\lambda-\epsilon\gamma_\mu D_\nu\bar\lambda)
 -\frac{1}{2\ell}(\bar\epsilon\gamma_{\mu\nu}\lambda
 +\epsilon\gamma_{\mu\nu}\bar\lambda)\,,\\
 \delta(D_\mu\sigma)&=\delta(\partial_\mu\sigma-i[A_\mu,\sigma])\nn\\
 &=\frac{1}{2}(\bar\epsilon D_\mu\lambda-\epsilon D_\mu\bar\lambda
 -\bar\epsilon\gamma_\mu[\lambda,\sigma]-\epsilon\gamma_\mu[\bar\lambda,\sigma])
 -\frac{i}{4\ell}(\bar\epsilon\gamma_\mu\lambda-\epsilon\gamma_\mu\bar\lambda)\,,\\
 \delta D+\delta\sigma/\ell&=
 -\frac{i}{2}\bar{\epsilon}(\gamma^{\mu}D_{\mu}\lambda-[\lambda,\sigma]+i\lambda/(2\ell))
 -\frac{i}{2}\epsilon(-\gamma^{\mu}D_{\mu}\bar{\lambda}-[\bar{\lambda},\sigma]
 -i\bar{\lambda}/(2\ell))\,,
\end{align}
%The terms of $\delta((\delta'\lambda)^{\dagger})\lambda$ which include 
%the covariant derivative: 
%\begin{align}
% \bar\epsilon'\epsilon(-i(D_\mu\bar\lambda)\gamma^\mu\lambda)
% +(\bar\epsilon'\gamma^\mu\bar\epsilon)(i(D_\mu\lambda)\lambda)\,,
%\end{align}
%The terms of $\delta((\delta'\lambda)^{\dagger})\lambda$ which include $\sigma$: 
%\begin{align}
% &\bar\epsilon'\epsilon(i[\bar\lambda,\sigma]\lambda) 
%  +\bar\epsilon'\bar\epsilon(i[\lambda,\sigma]\lambda) \nn\\
% &=\bar\epsilon'\epsilon(i\bar\lambda[\sigma,\lambda])\,,
%\end{align}
%where we used $\bar\epsilon'\bar\epsilon=0$
%and $[\bar\lambda,\sigma]\lambda=\bar\lambda[\sigma,\lambda]$ in the trace.\\
%The other terms of $\delta((\delta'\lambda)^{\dagger})\lambda$ : 
%\begin{align}
% &-\frac{1}{2\ell}\bar\epsilon'\epsilon(\bar\lambda\lambda)
% -\frac{1}{\ell}\bar\epsilon'\bar\epsilon(\lambda\lambda)\nn\\
% &=-\frac{1}{2\ell}\bar\epsilon'\epsilon(\bar\lambda\lambda)\,,
%\end{align}
%Putting these all together,
which are summed up to
\begin{align}
 \delta((\delta'\lambda)^{\dagger})\lambda= \bar\epsilon'\epsilon\Bigl(
 i\lambda\gamma^\mu D_\mu\bar\lambda+i\bar\lambda[\sigma,\lambda]
 -\frac{1}{2\ell}\bar\lambda\lambda
 \Bigr)+(\bar\epsilon'\gamma^\mu\bar\epsilon)(i(D_\mu\lambda)\lambda)\,.
\label{ll}
\end{align}
In (\ref{ll}), 
the last term is total derivative and it becomes a surface term:
\begin{align}
 i(\bar\epsilon'\gamma^\theta\bar\epsilon)(\lambda\lambda)|
 =i\bar\epsilon'\epsilon(\bar\lambda\gamma^\theta\lambda)|\,,
\end{align}
where we have used  \eqref{3d-bc-la} and \eqref{3d_e_be}.
Similarly, we find that
\begin{align}
 \bar\lambda\delta((\delta'\bar\lambda)^{\dagger})&=
 -\frac{1}{2}\bar\lambda\gamma^{\mu\nu}\bar\epsilon'\delta F_{\mu\nu}
 +i\bar\lambda\gamma^{\mu}\epsilon'\delta(D_\mu\sigma)
 -\bar\lambda\epsilon'(\delta D+\delta\sigma/\ell)\nn\\
 &=\bar\epsilon'\epsilon\Bigl(
 i\bar\lambda\gamma^\mu D_\mu\lambda
 +i\bar\lambda[\sigma,\lambda]
 -\frac{1}{2\ell}\bar\lambda\lambda
 \Bigr)
 -(\epsilon'\gamma^\mu\epsilon)(i(D_\mu\bar\lambda)\bar\lambda)\,,
\end{align}
where the last term is also 
a total derivative and it becomes a surface term:
\begin{align}
 i(\epsilon'\gamma^\theta\epsilon)(\bar\lambda\bar\lambda)|
 =i\bar\epsilon'\epsilon(\bar\lambda\gamma^\theta\lambda)|\,.
\end{align}
%where we used \eqref{3d_e_be} and \eqref{3d-bc-la}.
Thus, the fermionic part of $\delta$-exact term is
\begin{align}
 \delta((\delta'\lambda)^{\dagger})\lambda
 +\bar\lambda\delta((\delta'\bar\lambda)^{\dagger})
 &=\bar\epsilon'\epsilon\Bigl(
 i\bar\lambda\gamma^\mu D_\mu\lambda
 +i\lambda\gamma^\mu D_\mu\bar\lambda
 +2i\bar\lambda[\sigma,\lambda]
 -\frac{1}{\ell}\bar\lambda\lambda
 \Bigr)\\
 &=4\bar\epsilon'\epsilon\,\mathcal{L}_{\mathrm{YM}}^{\mathrm{fermion}}\,,
\end{align}
and there is no surface term.

Therefore, the $\delta$-exact Lagrangian 
$\delta V_{\mathrm{vector}} = \mathcal{L}_{\mathrm{YM}}^{\mathrm{boson}}+
\mathcal{L}_{\mathrm{YM}}^{\mathrm{fermion}} \,,$
is same as the Yang-Mills Lagrangian and 
there are no surface terms.
The saddle points of the bosonic part of $\delta V_{\mathrm{vector}}$
are given by
\begin{eqnarray}
 F_{\mu \nu}=0, \,\,\,\, D_\mu \sigma=0, \,\,\,\, D=-\frac{\sigma}{l},
\end{eqnarray}
which implies that
$A_\theta=0,\, A_{\tilde{\mu}}=a_{\tilde{\mu}},\, \sigma=\sigma_0$
where $a_{\tilde{\mu}}$ and $\sigma_0$ are constants satisfying 
$[ \sigma_0, a_{\tilde{\mu}} ] =0$ in an appropriate gauge.

\subsubsection{3D chiral multiplet}

For the chiral multiplet,
we consider the following $\delta$-exact term:
\begin{align}
\delta V_{\mathrm{chiral}} =
\frac{1}{2} 
\delta[(\delta'\psi)^\dagger\psi + \bar\psi(\delta'\bar\psi)^\dagger]
+\frac{q-1}{\ell}\,\delta[\bar\phi\,\delta'\phi-(\delta'\bar\phi)\phi]\,, 
\label{3d-deltaV-chiral}
\end{align}
where we  have defined
\begin{align}
(\delta'\psi)^\dagger &\equiv\bar\epsilon'\Bigl(-i D_\mu\bar\phi\gamma^\mu
-i \bar\phi\,\sigma-\frac{q}{\ell}\bar\phi\Bigr)-\epsilon' \bar F\,,
\\
(\delta'\bar\psi)^\dagger &\equiv\Bigl(-i D_\mu\phi\gamma^\mu
+i\sigma\,\phi+\frac{q}{\ell}\phi\Bigr)\epsilon'+\bar\epsilon' F\,.
\end{align}
The second term on the  right hand side of \eqref{3d-deltaV-chiral}
is added to simplify the calculation of the one-loop determinant.
We will compute the bosonic part of the $\delta$-exact term
first.
We can see that
\begin{align}
(\delta'\psi)^\dagger\delta\psi
&=\bar\epsilon'\epsilon\Bigl(
 D_\mu\bar\phi D^\mu\phi+\bar\phi\,\sigma^2\phi+\frac{q^2}{\ell^2}\,\bar\phi\,\phi+\bar F F
 \Bigr)
\nn\\
&\quad+\bar\epsilon'\gamma^\mu\epsilon\Bigl(
 i\varepsilon_{\mu\nu\rho}D^\nu\bar\phi D^\rho\phi
 +D_\mu\bar\phi\,\sigma\,\phi +i\frac{q}{\ell}D_\mu\bar\phi\,\phi
 +\bar\phi\,\sigma D_\mu\phi -i\frac{q}{\ell}\bar\phi D_\mu\phi
 \Bigr)
\nn\\ 
&\quad-i\epsilon'\gamma^\mu\epsilon\, \bar F D_\mu\phi
-i\bar\epsilon'\gamma^\mu\bar\epsilon\, D_\mu\bar\phi\, F\,,
\\%%%
\delta\bar\psi(\delta'\bar\psi)^\dagger
&= \bar\epsilon'\epsilon\Bigl(
 D_\mu\bar\phi D^\mu\phi+\bar\phi\,\sigma^2\phi+\frac{q^2}{\ell^2}\,\bar\phi\,\phi+\bar F F
 \Bigr)
\nn\\
&\quad +\bar\epsilon'\gamma^\mu\epsilon\Bigl(
 i\varepsilon_{\mu\nu\rho}D^\nu\bar\phi D^\rho\phi
 -D_\mu\bar\phi\,\sigma\,\phi +i\frac{q}{\ell}D_\mu\bar\phi\,\phi
 -\bar\phi\,\sigma D_\mu\phi -i\frac{q}{\ell}\bar\phi D_\mu\phi
 \Bigr)
\nn\\ 
&\quad +i\epsilon'\gamma^\mu\epsilon\, \bar F D_\mu\phi
+i\bar\epsilon'\gamma^\mu\bar\epsilon\, D_\mu\bar\phi\, F\,,
\end{align}
which give
\begin{align}
(\delta'\psi)^\dagger\delta\psi +\delta\bar\psi(\delta'\bar\psi)^\dagger
&=2\bar\epsilon'\epsilon\Bigl(
 D_\mu\bar\phi D^\mu\phi+\bar\phi\,\sigma^2\phi+\frac{q^2}{\ell^2}\,\bar\phi\,\phi+\bar F F
 \Bigr)
\nn\\
&\quad +2 i \,\bar\epsilon'\gamma^\mu\epsilon\Bigl(
 \varepsilon_{\mu\nu\rho}D^\nu\bar\phi D^\rho\phi
 +\frac{q}{\ell}D_\mu\bar\phi\,\phi
 -\frac{q}{\ell}\bar\phi D_\mu\phi
 \Bigr)\,.
\end{align}
We can also obtain
%\begin{align}
%\bar\phi\,(\delta\,\delta'\phi)&=
%\bar\epsilon'\epsilon\Bigl(i\bar\phi\,\sigma\phi-\frac{q}{\ell}\bar\phi\,\phi\Bigr)
%+i\bar\epsilon'\gamma^\mu\epsilon(\bar\phi D_\mu\phi)\,,
%\\%%%
%(\delta\,\delta'\bar\phi)\phi&=
%-\bar\epsilon'\epsilon\Bigl(i\bar\phi\,\sigma\phi-\frac{q}{\ell}\bar\phi\,\phi\Bigr)
%+i\bar\epsilon'\gamma^\mu\epsilon(D_\mu\bar\phi \, \phi)\,,
%\end{align}
%which give
\begin{align}
\bar\phi\,(\delta\,\delta'\phi)-(\delta\,\delta'\bar\phi)\phi &=
2\bar\epsilon'\epsilon\Bigl(i\bar\phi\,\sigma\phi-\frac{q}{\ell}\bar\phi\,\phi\Bigr)
+i\bar\epsilon'\gamma^\mu\epsilon(\bar\phi D_\mu\phi-D_\mu\bar\phi \, \phi)\,.
\end{align}

For the fermionic part of $\delta$-exact term,
we can see that
\begin{align}
\delta((\delta'\psi)^\dagger)\psi &=
\bar\epsilon'\epsilon\Bigl(i D_\mu\bar\psi\gamma^\mu\psi 
+i\bar\psi\,\sigma\,\psi-\frac{3 i}{2}\bar\phi\bar\lambda\psi
-\frac{1}{2\ell}\bar\psi\psi
\Bigr)
\nn\\
&\quad+\frac{i}{2}\bar\epsilon'\gamma^\mu\epsilon
\Bigl(2 i\varepsilon_{\mu\nu\rho}D^\nu\bar\psi\,\gamma^\rho\psi
+\bar\phi\bar\lambda\gamma_\mu\psi
-i\frac{2 q}{\ell}\bar\psi\gamma_\mu\psi
\Bigr)
\nn\\
&\quad+\frac{i}{2}\bar\epsilon'\gamma^\mu\bar\epsilon\,
\bar\phi\lambda\gamma_\mu\psi\,,
\\%%%
\bar\psi\delta((\delta'\bar\psi)^\dagger) &=
-\bar\epsilon'\epsilon\Bigl(i \bar\psi\gamma^\mu D_\mu\psi 
-i\bar\psi\,\sigma\,\psi-\frac{3 i}{2}\bar\psi\lambda\phi
+\frac{1}{2\ell}\bar\psi\psi
\Bigr)
\nn\\
&\quad -\frac{i}{2}\bar\epsilon'\gamma^\mu\epsilon
\Bigl(-2 i\varepsilon_{\mu\nu\rho}\bar\psi\,\gamma^\rho D^\nu\psi
+\bar\psi\gamma_\mu\lambda\phi
+i\frac{2 q}{\ell}\bar\psi\gamma_\mu\psi
\Bigr)
\nn\\
&\quad-\frac{i}{2}\epsilon'\gamma^\mu\epsilon\,
\bar\psi\gamma_\mu\bar\lambda\phi\,,
\end{align}
thus we find that
\begin{align}
&\delta((\delta'\psi)^\dagger)\psi +\bar\psi\delta((\delta'\bar\psi)^\dagger)
\nn\\
&=\bar\epsilon'\epsilon\Bigl(i D_\mu\bar\psi\gamma^\mu\psi -i \bar\psi\gamma^\mu D_\mu\psi
+2 i\bar\psi\,\sigma\,\psi-\frac{3 i}{2}\bar\phi\bar\lambda\psi +\frac{3 i}{2}\bar\psi\lambda\phi
-\frac{1}{\ell}\bar\psi\psi
\Bigr)
\nn\\
&\quad+\frac{i}{2}\bar\epsilon'\gamma^\mu\epsilon
\Bigl(2 i\varepsilon_{\mu\nu\rho}D^\nu\bar\psi\,\gamma^\rho\psi
+2 i\varepsilon_{\mu\nu\rho}\bar\psi\,\gamma^\rho D^\nu\psi
+\bar\phi\bar\lambda\gamma_\mu\psi -\bar\psi\gamma_\mu\lambda\phi
-i\frac{4 q}{\ell}\bar\psi\gamma_\mu\psi
\Bigr)
\nn\\
&\quad
-\frac{i}{2}\epsilon'\gamma^\mu\epsilon\,
\bar\psi\gamma_\mu\bar\lambda\phi
+\frac{i}{2}\bar\epsilon'\gamma^\mu\bar\epsilon\,
\bar\phi\lambda\gamma_\mu\psi\,.
\end{align}
We also compute
%\begin{align}
%\delta\bar\phi\,\delta'\phi &=
%-\frac{1}{2}\bar\epsilon'\epsilon\,\bar\psi\psi
%-\frac{1}{2}\bar\epsilon'\gamma^\mu\epsilon\,\bar\psi\gamma_\mu\psi\,,
%\\%%%
%\delta'\bar\phi\,\delta\phi &=
%\frac{1}{2}\bar\epsilon'\epsilon\,\bar\psi\psi
%+\frac{1}{2}\bar\epsilon'\gamma^\mu\epsilon\,\bar\psi\gamma_\mu\psi\,,
%\end{align}
%and then,
\begin{align}
&\delta\bar\phi\,\delta'\phi -\delta'\bar\phi\,\delta\phi =
-\bar\epsilon'\epsilon\,\bar\psi\psi
-\bar\epsilon'\gamma^\mu\epsilon\,\bar\psi\gamma_\mu\psi\,.
\end{align}
Therefore, the SUSY exact term we take is
\begin{align}
\delta V_{\mathrm{chiral}}
&= \bar\epsilon'\epsilon\Bigl(
 D_\mu\bar\phi D^\mu\phi+\bar\phi\,\sigma^2\phi +2 i\frac{q-1}{\ell}\bar\phi\,\sigma\,\phi
+\frac{q(2-q)}{\ell^2}\,\bar\phi\,\phi+\bar F F
\nn\\ 
&\qquad\qquad+\frac{i}{2} D_\mu\bar\psi\gamma^\mu\psi -\frac{i}{2} \bar\psi\gamma^\mu D_\mu\psi
+ i\bar\psi\,\sigma\,\psi-\frac{3 i}{4}\bar\phi\bar\lambda\psi +\frac{3 i}{4}\bar\psi\lambda\phi
-\frac{2 q-1}{2\ell}\bar\psi\psi
 \Bigr)
\nn\\
&\quad + i \,\bar\epsilon'\gamma^\mu\epsilon\Bigl(
 \varepsilon_{\mu\nu\rho}D^\nu\bar\phi D^\rho\phi
 +\frac{1}{\ell}D_\mu \bar\phi\,\phi
 -\frac{1}{\ell}\bar\phi D_\mu\phi
 +\frac{i}{2}\varepsilon_{\mu\nu\rho}D^\nu\bar\psi\,\gamma^\rho\psi
+\frac{i}{2}\varepsilon_{\mu\nu\rho}\bar\psi\,\gamma^\rho D^\nu\psi
\nn\\
&\qquad\qquad\qquad
+\frac{1}{4}\bar\phi\bar\lambda\gamma_\mu\psi -\frac{1}{4}\bar\psi\gamma_\mu\lambda\phi
-\frac{i}{\ell}\bar\psi\gamma_\mu\psi
 \Bigr)
 \nn\\
 &\quad
-\frac{i}{4}\epsilon'\gamma^\mu\epsilon\,
\bar\psi\gamma_\mu\bar\lambda\phi
+\frac{i}{4}\bar\epsilon'\gamma^\mu\bar\epsilon\,
\bar\phi\lambda\gamma_\mu\psi\,.
\end{align}
Note that there are no surface terms in this expression.
The bosonic part of $\int \delta V_{\mathrm{chiral}}$ can be written as
\begin{align}
 \int \rmd^3 x \sqrt{g}\,\left( \, 
\frac{1}{2} \left| D_\mu \phi 
-i \epsilon_{\mu \nu \rho} v^\nu D^\rho  \phi +\frac{i}{l} v_\mu \phi \right|^2
+\bar{\phi} (\sigma+i \frac{q-1}{l})^2 \phi
+\frac{1}{2} \left| \frac{1}{l} \phi-i v^\mu D_\mu \phi \right|^2
+\bar{F}F \right)\,,
\end{align}
which is positive definite after constant shift of $\sigma$ (and $D$).
Then, the saddle point of this is at $\phi=0,\, F=0$.
For other values of $\sigma$, it is reasonable to
think that the partition function and 
other exact quantities are obtained by the analytic continuation of $\sigma$. 
Alternatively, we can regard this $\delta V_{\mathrm{chiral}}$ as follows. 
If we use
$\delta[(\delta'\psi)^\dagger\psi +\bar\psi(\delta'\bar\psi)^\dagger]$,
which is manifestly positive definite,
as the $\delta$-exact terms for the matter multiplet,
we find that the saddle point equations are $\delta \psi=0$
and the saddle point 
is at $\phi=0,\, F=0$ partly because of the boundary conditions.
Because the addition of 
$\delta[\bar\phi\,\delta'\phi-(\delta'\bar\phi)\phi]$
will not change the 1-loop factor of the 
$\delta$-exact terms, we can think 
the saddle point of the $\int \delta V_{\mathrm{chiral}}$ for the matter multiplet
is at $\phi=0,\, F=0$ .

%%%%%%%%%%%%%%%%%%%%%%%%%%%%%%%%%%%%%%%%%%%%%%%%%%%%%%%%%%%%%
\subsection{One-loop determinant}
\label{3d-1loop}

\subsubsection{Vector multiplet}

We will compute the one-loop determinant
at the saddle point of $\int\delta V_{\mathrm{vector}}$.

With the addition of the term $t \delta V_{\mathrm{vector}}$,
we expand the fields around the saddle point as
\begin{align}
A_\theta &=a_\theta+\frac{1}{\sqrt{t}}\hat{A}_\theta
\,,\quad
A_{\tm} =a_{\tm}+\frac{1}{\sqrt{t}}\hat{A}_{\tm}
\,,\\
\sigma &=\sigma_0+\frac{1}{\sqrt{t}}\hat{\sigma}
\,,\quad
D =-\frac{\sigma_0}{\ell}+\frac{1}{\sqrt{t}}\hat{D}
\,,\\
\lambda &=\frac{1}{\sqrt{t}}\hat{\lambda}
\,,\quad
\bar\lambda =\frac{1}{\sqrt{t}}\hat{\bar\lambda}
\,.
\end{align}
where
$a_{\tm}$ and $\sigma_0$ are boundary values in 
\eqref{3d-bc-a} and \eqref{3d-bc-s},
and $a_\theta=0$.
We should set $a_\chi=0$ because of the smoothness at $\theta=0$.
On the other hand, 
$a_\varphi$ can be non-zero as an Wilson loop of 
the non-trivial 1-cycle of the manifold with the boundary
except for $\theta_0=\pi/2$.
%Hereafter, we assume that $a_\chi=0$ in the case where $\theta_0<\pi/2$
%and $a_\varphi=a_\chi=0$ in the case where $\theta_0=\pi/2$.

For the gauge fixing,
we will take the following Lorenz-like one:
\begin{align}
D^{(a)}_\mu \hat{A}^\mu=0\,,\quad(\da_\mu = \nabla_\mu -i[a_\mu\,,\cdot\,])\,.
\label{gauge_fixing}
\end{align}
The boundary conditions for the fluctuation fields
at $\theta=\theta_0$ are
\begin{align}
\hat{A}_{\tm}|=0\,,\quad
\hat{\sigma}|=0.
\label{3d_hat_bc}
\end{align}
With the gauge condition, we have
\begin{align}
t\int\rmd^3 x\sqrt{g} \mathcal{L}_{\mathrm{YM}}
=\int\rmd^3 x\sqrt{g} \,\tr\Bigl(
&-\frac{1}{2}A_\mu \da _\nu D^{(a)\nu} A^\mu
+\frac{1}{\ell^2}A_\mu A^\mu
-\frac{1}{2}[\sigma_0, \,A_\mu]^2
\nn\\
&-\frac{1}{2}\sigma\da_\mu D^{(a)\mu}\sigma
+(D+\sigma/\ell)^2
\nn\\
&+\frac{i}{4}\bar{\lambda}\gamma^{\mu}\da_\mu\lambda
+\frac{i}{4}\lambda\gamma^{\mu}\da_{\mu}\bar{\lambda}
+\frac{i}{2}\bar{\lambda}[\sigma_0,\lambda]-\frac{1}{4\ell}\bar{\lambda}\lambda
\Bigr)
\nn\\
&+\mathcal{O}(t^{-1/2})\,,
\label{lap1}
\end{align}
where we have omitted the hat symbols for the fluctuation fields,
for example $\hat{\sigma} \rightarrow \sigma$, for notational simplicity.
%and used the gauge fixing condition \eqref{gauge_fixing} 
%and boundary conditions \eqref{3d_hat_bc}.

Following the arguments in \cite{Hama:2011ea} 
for the squashed (ellipsoid) $S^3$, 
we only consider $A_\mu$ with the gauge conditions
and the fermions $\lambda, \bar{\lambda}$.
The (quadratic) kinetic operators for them are
\begin{eqnarray}
 L_2^A =\frac{1}{2} A_\mu \Delta_A A^\mu,
\end{eqnarray}
where
\begin{align}
\Delta_A &\equiv 
-\da _\nu D^{(a)\nu}+\frac{2}{\ell^2}+(\sa)^2 \,,
\end{align}
and
\begin{align}
L_2^{(\lambda,\bar{\lambda})} =\frac{1}{4}\tr
\begin{pmatrix}
\lambda &\bar\lambda
\end{pmatrix}
\begin{pmatrix}
0&\Delta_{\bar\lambda}\\
\Delta_\lambda &0
\end{pmatrix}
\begin{pmatrix}
\lambda \\
\bar\lambda
\end{pmatrix},
\end{align}
where
\begin{align}
\Delta_{\lambda} &\equiv 
i\gamma^\mu\da_\mu + i\sa-\frac{1}{2\ell}\,,\\
\Delta_{\bar\lambda} &\equiv 
i\gamma^\mu\da_\mu -i\sa-\frac{1}{2\ell}.
\end{align}
Note that 
we need to consider the eigenmodes for the pairs 
of $\lambda$ and $\bar{\lambda}$ because
the boundary conditions identify
$\lambda$ and $\bar{\lambda}$ at $\theta=\theta_0$,
thus they are not independent fields.

The 1-loop factor for the bosonic modes is given by $\prod_i M_i^{-1}$
up to an irrelevant numerical factor
where the eigenmode $A^\mu_i$ with the eigenvalue $M_i^2$ is 
defined by\footnote{
More precisely, the eigenvalue problem is 
\begin{eqnarray}
\int_M A'_\mu (\Delta_A -M_i^2) A^\mu_i=0,
\end{eqnarray}
where $A'_\mu$ is an arbitrary gauge field with the boundary condition
and the gauge condition.
Because $A'$ can take an arbitrary value with the gauge condition in 
the bulk, 
$\int_M A'_\mu A^\mu=0$ may mean $A^\mu=0$.
Thus, this eigenvalue problem may be same as (\ref{ep1}).
}
\begin{eqnarray}
\Delta_A A^\mu_i = M_i^2 A^\mu_i \,,\quad 
\mbox{with } \da_\mu A^\mu=0,
\label{ep1}
\end{eqnarray}
where $A^\mu_i$ should satisfy the boundary condition.
Note that $M_i^2$ is real and $M_i^2 > (\sa)^2$.

For the fermions, the 1-loop factor is given by
$\sqrt{\prod_i \nu_i}$ up to a numerical factor
where the eigenmode with the eigenvalue $\nu_i$ is given by\footnote{
The same argument as the bosonic modes can be applied for the fermions.}
\begin{align}
\begin{pmatrix}
0&C \Delta_{\bar\lambda}\\
C \Delta_\lambda &0
\end{pmatrix}
\begin{pmatrix}
\lambda_i \\
\bar\lambda_i
\end{pmatrix}
= \nu_i
\begin{pmatrix}
\lambda_i \\
\bar\lambda_i
\end{pmatrix},
\end{align}
which is the eigenvalue problem for the anti-symmetric 
operator although we will not use this operator.
Instead of this, we will consider the following 
eigenvalue problem:
\begin{align}
\begin{pmatrix}
0&\Delta_{\bar\lambda}\\
\Delta_\lambda &0
\end{pmatrix}
\begin{pmatrix}
\lambda_i \\
\bar\lambda_i
\end{pmatrix}
= \nu_i
\begin{pmatrix}
0&-1\\
1 &0
\end{pmatrix}
\begin{pmatrix}
\lambda_i \\
\bar\lambda_i
\end{pmatrix},
\end{align}
where the determinant of this operator and the original one 
are clearly same.
Thus the 1-loop factor
for the fermions is also given by the eigenvalues of this operator
as $\sqrt{\prod_i \nu_i}$, up to a numerical factor.
%We can see that 
%for this operator, 
%the vector $ P \begin{pmatrix}
%\lambda_i \\
%\bar\lambda_i 
%\end{pmatrix}
%\equiv 
%\begin{pmatrix}
%\bar\lambda_i^* \\
%\lambda_i^* 
%\end{pmatrix}$, 
%%where $I_G$ is an operator 
%%replacing $E_{\alpha}$ to 
%%$E_{-\alpha}$,
%has the eigenvalue $-\nu_i^*$.
%We can also see that
%$\nu_i^* (\alpha) = \nu_i (-\alpha)$
%where $\nu_i (\alpha)$ is an eigenvalue of an eigenmode
%proportional to $E_\alpha$ which is the base for the root $\alpha$.

Now, basically following \cite{Hama:2011ea},
we will show that almost all the eigenvalues in the bosonic and the fermionic 
1-loop factors are canceled.
Let us suppose $A_\mu$ be a bosonic eigenmode,
\begin{align}
\Delta_A A_\mu = M^2 A_\mu \,,\quad 
\da_\mu A^\mu=0\,.
\end{align}
Then, the following two $\lambda$ modes
\begin{align}
\lambda_1\equiv \gamma^\mu \epsilon A_\mu\,,
\quad
\lambda_2 \equiv -\varepsilon^{\mu\nu\rho}\gamma_\rho\,\epsilon\,\da_\mu A_\nu \,,
\end{align}
satisfy
\begin{align}
\Delta_\lambda
\begin{pmatrix}
\lambda_1\\
\lambda_2
\end{pmatrix}
=
\begin{pmatrix}
i\sa &1\\
M^2 -(\sa)^2 & i\sa
\end{pmatrix}
\begin{pmatrix}
\lambda_1\\
\lambda_2
\end{pmatrix}
\,.
\end{align}
Then, the following linear combinations of them,
\begin{align}
\lambda_{\pm}\equiv \pm \sqrt{M^2-(\sa)^2}\,\lambda_1+\lambda_2 \,,
\end{align}
satisfy
\begin{align}
\Delta_\lambda \lambda_{\pm} = \nu_{\pm} \lambda_{\pm}
\quad\quad
\Bigl( \nu_{\pm}= i\sa\pm \sqrt{M^2-(\sa)^2} \Bigr) \,.
\end{align}
Similarly, the following two $\bar{\lambda}$ modes
\begin{align}
\bar{\lambda}_1\equiv -\gamma^\mu \bar\epsilon A_\mu\,,
\quad
\bar{\lambda}_2 \equiv -\varepsilon^{\mu\nu\rho}\gamma_\rho\,\bar\epsilon\,\da_\mu A_\nu \,,
\end{align}
satisfy
\begin{align}
\Delta_{\bar\lambda}
\begin{pmatrix}
\bar\lambda_1\\
\bar\lambda_2
\end{pmatrix}
=
\begin{pmatrix}
-i\sa &-1\\
-M^2 +(\sa)^2 & -i\sa
\end{pmatrix}
\begin{pmatrix}
\bar{\lambda}_1\\
\bar{\lambda}_2
\end{pmatrix}
\,,
\end{align}
and the linear combinations of them
\begin{align}
\bar{\lambda}_{\pm}\equiv \pm \sqrt{M^2-(\sa)^2}\,\bar{\lambda}_1+\bar{\lambda}_2 \,,
\end{align}
satisfy
\begin{align}
\Delta_{\bar\lambda} \bar{\lambda}_{\pm} = \bar{\nu}_{\pm} \bar{\lambda}_{\pm}
\quad\quad
\Bigl( \bar{\nu}_{\pm}= -i\sa\mp  \sqrt{M^2-(\sa)^2}  = -\nu_\pm \Bigr) \,.
\end{align}

Therefore, we have two fermionic eigenmodes:
$\begin{pmatrix}
\lambda_+\\
\bar{\lambda}_+
\end{pmatrix}$
and
$\begin{pmatrix}
\lambda_-\\
\bar{\lambda}_-
\end{pmatrix}$
whose eigenvalues are $\nu_+$ and $\nu_-=-\nu_+^*$.
We can show that 
these modes satisfy the boundary condition 
\begin{align}
\ell e^{i(\varphi-\chi)}\gamma^\theta \lambda_{\pm}| =\bar{\lambda}_{\pm}| \,,
\end{align}
where we have used the fact that 
$A_\mu$ satisfies the boundary condition ($A_{\tilde{\mu}} |=0$).
The contribution of these two fermion modes
is canceled by the corresponding bosonic eigenmode 
because $\sqrt{-\nu_+ \nu_-}=M$.

Conversely, let us suppose 
that $\begin{pmatrix}
\lambda\\
\bar{\lambda}
\end{pmatrix}$
be a fermionic eigenmode with,
\begin{align}
\Delta_\lambda \lambda= \nu \lambda
\,,\quad
\Delta_{\bar\lambda} \bar\lambda = -\nu \bar\lambda 
\,,
\end{align}
which satisfies the boundary condition
$ \ell e^{i(\varphi-\chi)}\gamma^\theta \lambda| =\bar{\lambda}| \,$.
Then we can construct a bosonic eigenmode
\begin{align}
 A_\mu \equiv
 (\nu-i\sa)
(\bar\epsilon \gamma_\mu\lambda+\epsilon\gamma_\mu\bar\lambda)
-i\da_\mu(\bar\epsilon\lambda-\epsilon\bar\lambda)
\,,
\end{align}
which satisfies the boundary condition ($ A_{\tilde{\mu}} | =0$)
and the gauge fixing condition ($\da_\mu A^\mu=0$).
The corresponding eigenvalue is ${M'}^2=(\nu-i\sa)^2+(\sa)^2$, i.e.
\begin{align}
\Delta_A A_\mu = [(\nu-i\sa)^2+(\sa)^2]A_\mu\,.
\end{align}
Because ${M'}^2$ should be real and $M^2 > (\sa)^2$,
we have $\nu= i \sa \pm \sqrt{{M'}^2-(\sa)^2}$ and 
$-\nu^*= i \sa \mp \sqrt{{M'}^2-(\sa)^2}$.
Therefore the 1-loop contributions of
$\begin{pmatrix}
\lambda\\
\bar{\lambda}
\end{pmatrix}$
and
the fermionic mode corresponding the eigenvalue $-\nu^*$
%$P \begin{pmatrix}
%\lambda\\
%\bar{\lambda}
%\end{pmatrix}$ 
are canceled by the one from the bosonic eigenmode.

\paragraph{Unpaired eigenmodes}~\\

%XXXX In this paragraph,
%we assume $\epsilon$ and $\bar\epsilon$ as Grassmann-even. 
%Is this remark needed? XXX
We have shown that there is a map between 
the bosonic and fermionic eigenmodes
and almost all the eigenvalues are canceled in evaluating the 1-loop factor.
Hence, we may study only the bosonic (fermionic) eigenmodes
whose eigenvalues are not canceled by the fermionic (bosonic) eigenvalues.
We will call such modes as  
the unpaired bosonic (fermionic) eigenmodes.
In Appendix \ref{useful_3d}\,, we summarize
some useful formulas for Killing spinors and Killing vectors 
\eqref{v}, \eqref{vplus}, \eqref{vminus}, 
which will be used in the following calculation.

The unpaired bosonic eigenmodes 
should satisfy 
$\begin{pmatrix}
\lambda_+\\
\bar{\lambda}_+
\end{pmatrix}=0$
or 
$\begin{pmatrix}
\lambda_-\\
\bar{\lambda}_-
\end{pmatrix}=0$,
where $\lambda_\pm, \bar\lambda_\pm$ are the corresponding 
fermionic modes, as we can see from the discussions of the cancellations.
%We can expand $A_\mu$ as
Any unpaired bosonic mode
can be written using the general decomposition of the gauge field:
\begin{align}
A_\mu \equiv v_\mu Y+v^+_\mu Y_+ +v^-_\mu Y_- =v^X_\mu Y_X\,,
\end{align}
where $v^X_\mu$ ($X = \text{no mark},\,+,\,-$) are defined by \eqref{v}-\eqref{vminus}.
The boundary conditions can be rewritten into
\begin{align}
Y|=0\,,\quad
(e^{-i(\varphi-\chi)}Y_+ +e^{i(\varphi-\chi)}Y_-)|=0\,,
\label{bc1}
\end{align}
and the gauge fixing condition, $\da_\mu A^\mu =0$, becomes
\begin{align}
 v_X^\mu\da_\mu Y_X=0
 \,,
 \label{gaugefixY}
\end{align}
where we have used
$ \da_\mu A_\nu = v^X_\nu\da_\mu Y_X 
+\frac{1}{\ell}\varepsilon_{\mu\nu\rho}v_X^\rho Y_X\,Y$.
The eigenvalue equation, $\Delta_A A_\mu = M^2 A_\mu$, can be written as
\begin{align}
-v^X_\mu D^{(a)\nu}\da_\nu Y_X 
+\frac{2}{\ell}\varepsilon_{\mu\nu\rho}v_X^\rho D^{(a)\nu}Y_X
=\biggl[M^2-(\sa)^2-\frac{4}{\ell^2}\biggr]v^X_\mu Y_X \,,
\end{align}
which can be decomposed into the following three equations:
\begin{align}
-D^{(a)\nu}\da_\nu Y 
+\frac{2i}{\ell}v_+^\nu \da_\nu Y_+
-\frac{2i}{\ell}v_-^\nu \da_\nu Y_-
&=\biggl[M^2-(\sa)^2-\frac{4}{\ell^2}\biggr]Y
\,,
\label{eigenY}
\\
-D^{(a)\nu}\da_\nu Y_- 
+\frac{i}{\ell}v_+^\nu \da_\nu Y
+\frac{2i}{\ell}v^\nu \da_\nu Y_-
&=\biggl[M^2-(\sa)^2-\frac{4}{\ell^2}\biggr]Y_-
\,,
\label{eigenY-}\\
-D^{(a)\nu}\da_\nu Y_+ 
-\frac{i}{\ell}v_-^\nu \da_\nu Y
-\frac{2i}{\ell}v^\nu \da_\nu Y_+
&=\biggl[M^2-(\sa)^2-\frac{4}{\ell^2}\biggr]Y_+
\,.
\label{eigenY+}
\end{align}

The unpaired bosonic modes should satisfy
$C \lambda_1+\lambda_2 =0$,
more explicitly, 
$
\Bigl(C- \frac{2}{\ell}\Bigr)(Y\epsilon +2Y_-\bar\epsilon)
+i(\gamma^\mu\epsilon\da_\mu Y +2\gamma^\mu\bar\epsilon\da_\mu Y_-)
=0$
which is equivalent to
\begin{align}
-2\Bigl(C- \frac{2}{\ell}\Bigr)Y_- 
 +iv^\mu_+\da_\mu Y +2i v^\mu\da_\mu Y_-
&=0\,,
\label{Y_-Y1}
\\
\Bigl(C- \frac{2}{\ell}\Bigr)Y 
 +iv^\mu\da_\mu Y +2i v^\mu_-\da_\mu Y_-
&=0\,,
\label{Y_-Y2}
\end{align}
where
we have used
\begin{align}
\lambda_1 &=\gamma^\mu \epsilon A_\mu
 =Y\epsilon +2Y_-\bar\epsilon
\,,\\
\lambda_2 &=-\varepsilon^{\mu\nu\rho}\gamma_\rho\,\epsilon\,\da_\mu A_\nu
 =-\frac{2}{\ell}(Y\epsilon +2Y_-\bar\epsilon) 
  +i(\gamma^\mu\epsilon\da_\mu Y +2\gamma^\mu\bar\epsilon\da_\mu Y_-)
\,,\\
\bar\lambda_1 &=-\gamma^\mu \bar\epsilon A_\mu
 =Y\bar\epsilon +2Y_+\epsilon
\,,\\
\bar\lambda_2 &=-\varepsilon^{\mu\nu\rho}\gamma_\rho\,\bar\epsilon\,\da_\mu A_\nu
 =\frac{2}{\ell}(Y\bar\epsilon +2Y_+\epsilon) 
  -i(\gamma^\mu\bar\epsilon\da_\mu Y +2\gamma^\mu\epsilon\da_\mu Y_+)
\,.
\end{align}
and defined
\begin{eqnarray}
C \equiv \pm\sqrt{M^2-(\sa)^2}\,.
\end{eqnarray}
For the other condition, $C \bar\lambda_1+\bar\lambda_2 =0$\,,
we have 
\begin{align}
\Bigl(C+ \frac{2}{\ell}\Bigr)(Y\bar\epsilon +2Y_+\epsilon)
-i(\gamma^\mu\bar\epsilon\da_\mu Y +2\gamma^\mu\epsilon\da_\mu Y_+)
=0\,,
\end{align}
which is equivalent to
\begin{align}
2\Bigl(C+ \frac{2}{\ell}\Bigr)Y_+ 
 -iv^\mu_-\da_\mu Y -2i v^\mu\da_\mu Y_+
&=0\,,
\label{Y_+Y1}
\\
-\Bigl(C+ \frac{2}{\ell}\Bigr)Y 
 -iv^\mu\da_\mu Y -2i v^\mu_+\da_\mu Y_+
&=0\,.
\label{Y_+Y2}
\end{align}

Now we found all the equations for the unpaired bosonic modes
and will solve them.
From \eqref{gaugefixY}, \eqref{Y_-Y2} and \eqref{Y_+Y2},
we obtain
\begin{align}
C\, Y = 0\,.
\end{align}
We can easily see that there are no nontrivial solutions if $C=0$, 
which implies $\lambda_2=\bar\lambda_2=0$ and then 
$\varepsilon^{\mu \nu \rho} \da_\mu A_\nu=0$.
Therefore, we will solve them for $Y=0$.
This implies 
\begin{align}
\Bigl(C- \frac{2}{\ell}\Bigr)Y_- 
-i v^\mu\da_\mu Y_-
&=0\,,\quad
v^\mu_-\da_\mu Y_-
=0\,,
\label{upm1}
\end{align}
and
\begin{align}
\Bigl(C+ \frac{2}{\ell}\Bigr)Y_+ 
 -i v^\mu\da_\mu Y_+
&=0\,,\quad
v^\mu_+\da_\mu Y_+
=0\,.
\label{upp1}
\end{align}
Using the boundary conditions, we take the following ansatz:
\begin{align}
Y_-=f_- (\theta)e^{i ((m-1)\varphi -(n-1)\chi)}
\,,\quad
Y_+=f_+ (\theta)e^{i ((m+1) \varphi -(n+1) \chi)} \,,
\end{align}
where $m$ and $n$ are integers.
Then, the $v^\mu_\mp \da_\mu Y_\mp=0$ is solved as
\begin{align}
 f_-(\theta)\propto \cos^{m-1-a_\varphi}\theta \sin^{n-1}\theta
 \,,\quad
 f_+(\theta)\propto \cos^{-m-1+a_\varphi}\theta\sin^{-n-1}\theta
 \,.
\end{align}
Because of the regularity at $\theta=0$,
we have two cases: $f_+=0, n \geq 1$
or $f_-=0, n \leq -1$, however,
the boundary conditions (\ref{bc1})
do not allow both cases 
for $\theta_0<\pi/2$.

Therefore, we conclude that
there is no bosonic unpaired eigenmode for $\theta_0<\pi/2$.
For the special case that $\theta_0=\pi/2$,
there are the following unpaired bosonic eigenmodes:
\begin{align}
 Y&=Y_+=0\,,\quad
 Y_-\propto\cos^{m-1}\theta\,\sin^{n-1}\theta\, e^{i(m-1)\varphi-i(n-1)\chi}\,,\nn\\
 C&=\frac{m+n}{\ell}>0\quad(m\geq2\,,\,\, n\geq1)\,,
\end{align}
or
\begin{align}
 Y&=Y_-=0\,,\quad
 Y_+\propto\cos^{-m-1}\theta\,\sin^{-n-1}\theta\, e^{i(m+1)\varphi-i(n+1)\chi}\,,\nn\\
 C&=\frac{m+n}{\ell}<0\quad(m\leq-2\,,\,\, n\leq-1)\,.
\end{align}
For $S^3$, i.e. the case without boundary, 
the unpaired bosonic eigenmodes
are given by
\begin{align}
 Y&=Y_+=0\,,\quad
 Y_-\propto\cos^{m-1}\theta\,\sin^{n-1}\theta\, e^{i(m-1)\varphi-i(n-1)\chi}\,,\nn\\
 C&=\frac{m+n}{\ell}>0\quad(m\geq1\,,\,\, n\geq1)\,,
\end{align}
or
\begin{align}
 Y&=Y_-=0\,,\quad
 Y_+\propto\cos^{-m-1}\theta\,\sin^{-n-1}\theta\, e^{i(m+1)\varphi-i(n+1)\chi}\,,\nn\\
 C&=\frac{m+n}{\ell}<0\quad(m\leq-1\,,\,\, n\leq-1)\,,
\end{align}

%%%%%%%%%%%%%%%%%%%%%%%%%%%%%%%%%%%%%%%%%%%%%%%%%%%%%%%%%%%%%
Let us consider unpaired fermionic eigenmodes,
\begin{align}
\Delta_\lambda \lambda= \nu \lambda
\,,\quad
\Delta_{\bar\lambda} \bar\lambda = -\nu \bar\lambda 
\,,
\end{align}
for which the corresponding bosonic eigenmode vanishes, i.e. 
\begin{align}
 (\nu-i\sa)
(\bar\epsilon \gamma_\mu\lambda+\epsilon\gamma_\mu\bar\lambda)
-i\da_\mu(\bar\epsilon\lambda-\epsilon\bar\lambda)=0
\,.
\label{missing vector}
\end{align}
We can expand $\lambda$ and $\bar\lambda$ as
\begin{align}
\lambda =\Lambda\epsilon +\Lambda'\bar\epsilon
\,,\quad 
\bar\lambda =\bar\Lambda\bar\epsilon +\bar\Lambda'\epsilon
\,,
\end{align}
where $\Lambda,\Lambda',\bar\Lambda$ and $\bar\Lambda'$ are scalars.
Then the boundary conditions can be written as
\begin{align}
\Lambda|+\bar\Lambda| =0
\,,\quad
e^{i(\varphi-\chi)}\Lambda'|+e^{-i(\varphi-\chi)}\bar\Lambda'| =0\,.
\end{align}
In this expansion,
the eigenvalue equation, $\Delta_\lambda \lambda= \nu \lambda$, is equivalent to
\begin{align}
i v_+^\mu\da_\mu \Lambda +i v^\mu\da_\mu \Lambda' 
&=-\Bigl(\nu -i\sa+\frac{2}{\ell}\Bigr) \Lambda'\,,
\label{eigen lambda1}\\
i v^\mu\da_\mu \Lambda +i v_-^\mu\da_\mu \Lambda' 
&=\Bigl(\nu -i\sa+\frac{2}{\ell}\Bigr) \Lambda\,,
\label{eigen lambda2}
\end{align}
and $\Delta_{\bar\lambda} \bar\lambda= -\nu \bar\lambda$ is equivalent to
\begin{align}
 i v^\mu\da_\mu \bar\Lambda +i v_+^\mu\da_\mu \bar\Lambda' 
&=\Bigl(\nu -i\sa-\frac{2}{\ell}\Bigr) \bar\Lambda\,,
\label{eigen barlambda1}\\
 i v_-^\mu\da_\mu \bar\Lambda +i v^\mu\da_\mu \bar\Lambda' 
&=-\Bigl(\nu -i\sa-\frac{2}{\ell}\Bigr) \bar\Lambda'\,.
\label{eigen barlambda2}
\end{align}
The equation \eqref{missing vector} 
can be put 
into the form
\begin{align}
i v^\mu \da_\mu (\Lambda+\bar\Lambda) &=(\nu-i\sa)(\Lambda+\bar\Lambda) \,,
\label{vA=0}\\
i v_+^\mu \da_\mu (\Lambda+\bar\Lambda) &=-2(\nu-i\sa)\Lambda' \,, 
\label{v+A=0}\\
i v_-^\mu \da_\mu (\Lambda+\bar\Lambda) &=-2(\nu-i\sa)\bar\Lambda' \,.
\label{v-A=0}
\end{align}

We will solve the equations above for the unpaired fermionic modes.
By \eqref{eigen lambda1}, \eqref{vA=0} and \eqref{v+A=0},
we obtain
\begin{align}
 v_+^\mu\da_\mu \bar\Lambda =0\,,
 \label{v+barlambda=0}
\end{align}
and by \eqref{eigen barlambda2}, \eqref{vA=0} and \eqref{v-A=0},
we obtain
\begin{align}
v_-^\mu\da_\mu \Lambda =0\,. 
\label{v-lambda=0}
\end{align}
If we make the ansatz,
\begin{align}
\Lambda=f(\theta)e^{i (m\varphi -n\chi)}
\,,\quad
\bar\Lambda=\bar{f}(\theta)e^{i (m\varphi -n\chi)} \,,
\end{align}
where $m$ and $n$ are integers,
then we can solve \eqref{v+barlambda=0} and \eqref{v-lambda=0} as
\begin{align}
 f(\theta)\propto \cos^{m-a_\varphi}\theta \sin^{n}\theta
 \,,\quad
 \bar{f}(\theta)\propto \cos^{-m+a_\varphi}\theta\sin^{-n}\theta
 \,.
\end{align}

For $\theta_0<\pi/2$,
the boundary conditions and the regularity at $\theta=0$ 
fix $f, \bar{f}$ as
\begin{align}
f(\theta)=f_0\biggl(\frac{\cos\theta}{\cos\theta_0}\biggr)^{m-a_\varphi} 
\,,\quad
\bar{f}(\theta)=-f_0\biggl(\frac{\cos\theta}{\cos\theta_0}\biggr)^{-m+a_\varphi}\,,
\end{align}
where $f_0$ is a constant.
Then eq. \eqref{vA=0} determines the eigenvalue as
\begin{align}
 \nu= i\sa+\frac{m-a_\varphi}{\ell}\,.
\end{align}
Furthermore, from \eqref{v+A=0} and \eqref{v-A=0}, 
$\Lambda', \bar\Lambda'$ should take the following forms
\begin{align}
 \Lambda'&=i f_0\frac{\sin\theta}{\cos\theta}
\biggl(\frac{\cos\theta}{\cos\theta_0}\biggr)^{m-a_\varphi}
e^{i(m-1)\varphi +i \chi} \,,
\\
 \bar\Lambda'&= -i f_0\frac{\sin\theta}{\cos\theta}
\biggl(\frac{\cos\theta}{\cos\theta_0}\biggr)^{-m+a_\varphi}
e^{i(m+1)\varphi -i \chi} \,,
\end{align}
which are consistent with the boundary conditions.
We can also check that these satisfy
\eqref{eigen lambda2} and \eqref{eigen barlambda1}.
Thus, these solutions are indeed the 
fermionic unpaired eigenmodes for $\theta_0<\pi/2$.

For $\theta_0=\pi/2$,\footnote{
Note that $a_\varphi=a_\chi=0$ in this case.}
%we do not have above solutions since they are singular.
%However, we have other solutions
we find that the followings are the consistent solutions:
\begin{align}
\Lambda&=f_0\,\cos^m\theta\sin^n\theta \, e^{i (m\varphi -n\chi)}
\,,\quad
\bar\Lambda=0\,,
\\
\Lambda'&=\frac{i f_0}{m+n}(m\cos^{m-1}\theta\sin^{n+1}\theta
-n\cos^{m+1}\theta\sin^{n-1}\theta) e^{i (m-1)\varphi -i(n-1)\chi}
\,,\quad
\bar\Lambda'=0\,,
\\
\nu &=i\sa+\frac{m+n}{\ell}
\qquad
(m\geq 2\,,\, n\geq 1)\,,
\end{align}
and
\begin{align}
\Lambda&=0
\,,\quad
\bar\Lambda=f_0\,\cos^{-m}\theta\sin^{-n}\theta \, e^{i (m\varphi -n\chi)}\,,
\\
\Lambda'&=0
\,,\quad
\bar\Lambda'=\frac{i f_0}{m+n}(m\cos^{-m-1}\theta\sin^{-n+1}\theta
-n\cos^{-m+1}\theta\sin^{-n-1}\theta) e^{i (m+1)\varphi -i(n+1)\chi}\,,
\\
\nu &=i\sa+\frac{m+n}{\ell}
\qquad
(m\leq -2\,,\, n\leq -1)\,.
\end{align}

For $S^3$,
the fermionic unpaired eigenmodes 
are given by
\begin{align}
\Lambda&=f_0\,\cos^m\theta\sin^n\theta \, e^{i (m\varphi -n\chi)}
\,,\quad
\bar\Lambda=0\,,
\\
\Lambda'&=\frac{i f_0}{m+n}(m\cos^{m-1}\theta\sin^{n+1}\theta
-n\cos^{m+1}\theta\sin^{n-1}\theta) e^{i (m-1)\varphi -i(n-1)\chi}
\,,\quad
\bar\Lambda'=0\,,
\\
\nu &=i\sa+\frac{m+n}{\ell}
\qquad
(m\geq 0\,,\, n\geq 0\,,\, m n \neq 0)\,,
\end{align}
and
\begin{align}
\Lambda&=0
\,,\quad
\bar\Lambda=f_0\,\cos^{-m}\theta\sin^{-n}\theta \, e^{i (m\varphi -n\chi)}\,,
\\
\Lambda'&=0
\,,\quad
\bar\Lambda'=\frac{i f_0}{m+n}(m\cos^{-m-1}\theta\sin^{-n+1}\theta
-n\cos^{-m+1}\theta\sin^{-n-1}\theta) e^{i (m+1)\varphi -i(n+1)\chi}\,,
\\
\nu &=i\sa+\frac{m+n}{\ell}
\qquad
(m\leq 0\,,\, n\leq 0\,,\, m n \neq 0)\,.
\end{align}

Therefore, up to an overall numerical factor,\footnote{
We have also neglected a factor like $\ell^{p}$ where $p$ is a number.} 
%which is independent of the boundary values (e.g. $\sigma_0$),
the 1-loop factor for the vector multiplet for $\theta_0<\pi/2$ is given by
\begin{align}
 Z^{\rm{1-loop}}_{\rm{vector}} &=
 \prod_{\alpha\in\Delta_+}
 \prod_{m}\, (i\alpha(\sigma_0)\,\ell +m-\alpha(a_\varphi))\,,
\end{align}
where $\Delta_+$ is the set of the positive roots.
For the special value $\theta_0=\pi/2$\,,
the 1-loop factor is
\begin{align}
 Z^{\rm{1-loop}}_{\rm{vector}} &=
 \prod_{\alpha\in\Delta_+}
 \frac{\prod_{|m|\geq 2\,, |n|\geq 1\,, mn>0}\, (i\alpha(\sigma_0)\,\ell -m-n)\,(i\alpha(\sigma_0)\,\ell +m+n)}
{\prod_{|m|\geq 2\,,|n|\geq 1\,, mn>0}(\alpha(\sigma_0)^2\ell^2+(m+n)^2)}
\nn\\
&=1 %\text{const.}
\,.
\end{align}

For $S^3$\,, the 1-loop factor is given by
\begin{align}
Z^{\rm{1-loop}}_{\rm{vector}} &=
 \prod_{\alpha\in\Delta_+}
 \frac{\prod_{|m|\geq 1\,, |n|\geq 1\,, mn>0}\, (i\alpha(\sigma_0)\,\ell -m-n)
        \prod_{|m|\geq 0\,, |n|\geq 0\,, mn>0}\, (i\alpha(\sigma_0)\,\ell +m+n)}
{\prod_{|m|\geq 1\,,|n|\geq 1\,, mn>0}(\alpha(\sigma_0)^2\ell^2+(m+n)^2)}
\nn\\
&=\prod_{\alpha\in\Delta_+}
 \frac{1}{\alpha(\sigma_0)^2\ell^2}
 \prod_{m\geq 0\,, n \geq 0}
 \frac{(\alpha(\sigma_0)^2\ell^2 +(m+n)^2)}
{(\alpha(\sigma_0)^2\ell^2+(m+n+2)^2)}
\nn\\
&=\prod_{\alpha\in\Delta_+}
 \prod_{m\geq 1}
 (\alpha(\sigma_0)^2\ell^2 +m^2)^2 \,,
\end{align}
which, of course, coincides with the one
obtained in \cite{Kapustin:2009kz, Hama:2011ea}.

%%%%%%%%%%%%%%%%%%%%%%%%%%%%%%%%%%%%%%%%%%%%%%%%%%%%%%%%%%%%%
\subsubsection{Chiral multiplet}

Next, let us consider the chiral multiplet.
Expanding the fields around the saddle point
and leaving only the quadratic terms, 
we have
\begin{align}
t \int \rmd^3 x \sqrt{g}\, \delta V_{\mathrm{chiral}} &=
 \int \rmd^3 x \sqrt{g}\, \mathcal{L}_{\text{reg}}
 +\mathcal{O}(t^{-1/2})\,,
\end{align}
where
\begin{align}
 \mathcal{L}_{\text{reg}} &=
\bar\phi\,
\Delta_\phi\,
\phi 
+\bar\psi\,
\Delta_\psi\,
\psi
\,,
\\
\Delta_\phi &=
-\da_\mu D^{(a)\mu} +\sigma_0^2 +2 i\frac{q-1}{\ell}\sigma_0+\frac{q(2-q)}{\ell^2}\,,
\\
\Delta_\psi &=
-i\gamma^\mu\da_\mu +i\sigma_0 -\frac{2 q-1}{2\ell}\,.
\end{align}
Hereafter, we set 
\begin{eqnarray}
 \omega = i\sigma_0 -(q-1)/\ell \,.
\end{eqnarray}
The eigenvalue problems for the 1-loop factor
are $\Delta_\phi \phi=M^2 \phi$ and
$\Delta_\psi\psi=\nu\,\psi$.
Note that the boundary conditions for  
$\psi$ and $\bar\psi$ are independently imposed, thus
we can consider the eigenvalue problems independently.
For $\bar\psi$, we have $\Delta_\psi \bar\psi=\nu\, \bar\psi$
where $\sigma_0$ in $\Delta_\psi$ are in the complex conjugate
representation of the one for $\psi$. 
%Thus, we can construct the eigenmode $\bar\psi$ with the eigenvalue $\nu$
%for the eigenmode $\psi$ with the eigenvalue $\nu$.

The cancellations in the 1-loop factor 
between the contributions from the bosonic and fermionic modes
can be seen as follows.\footnote{
The cancellations in the 1-loop factor can be seen 
more transparently as we will show in Appendix {\ref{can1}}.}

Let us suppose $\psi$ as a fermionic eigenmode:
$\Delta_\psi\psi=\nu\,\psi$ satisfying the boundary conditions
\eqref{3d_chiral_bc}.
Then, if we define $\phi_1\equiv \bar\epsilon\psi$,
we find that $\phi_1$ is a scalar eigenmode, i.e. 
$\Delta_\phi \phi_1=\nu(\nu-2\omega)\,\phi_1$,
which satisfies the the boundary condition.

Conversely, for a given scalar eigenmode
($\Delta_\phi \phi =M^2\phi$) satisfying the boundary condition,
we define
\begin{align}
\psi_\pm \equiv \Bigl(\nu_\pm-\omega+\frac{1}{\ell}\Bigr)\,\epsilon\phi
-i\gamma^\mu \epsilon\, \da_\mu\phi \,,
\end{align}
where $\nu_\pm\equiv\omega\pm\sqrt{M^2+\omega^2}$\,.
Then, we find that
\begin{align}
\Delta_\psi \psi_\pm =
\nu_\pm\,\psi_\pm
\,,
\end{align}
and $\psi_\pm$ satisfy the boundary condition.

%Note that $\phi_1$ and $\psi_\pm$ satisfy the boundary condition
%\eqref{3d_chiral_bc}.
Therefore, we may evaluate only the eigenvalues 
which are not canceled.

\paragraph{Unpaired eigenmodes}~\\

The unpaired fermionic eigenmode $\psi$  should satisfy
$\phi_1(\equiv \bar\epsilon \psi)=0$.
For such modes, 
we can take $\psi=\bar\epsilon\Psi$,
where $\Psi$ is a scalar function on which any boundary condition is
not imposed. 
Since $\psi$ is a fermionic eigenmode, $\Delta_\psi\psi=\nu\,\psi$,
we have 
\begin{align}
v_-^\mu\da_\mu \Psi &=0\,,\\
i v^\mu\da_\mu\Psi &=\Bigl(\nu-\omega-\frac{1}{\ell}\Bigr)\Psi\,.
\end{align}
Therefore, we obtain
\begin{align}
 \Psi &\propto\cos^{m-a_\varphi}\theta\,\sin^n\theta\,
 e^{i m\varphi-i n\chi}\,,\\
 \nu &=i\sigma_0-\frac{q-2}{\ell}+\frac{m-a_\varphi+n}{\ell}
\quad (m\in \mathbb{Z} \,,\, n\geq 0)\,,
\end{align}
for $\theta_0<\pi/2$\,,
or
\begin{align}
 \Psi &\propto\cos^{m}\theta\,\sin^n\theta\,
 e^{i m\varphi-i n\chi}\,,\\
 \nu &=i\sigma_0-\frac{q-2}{\ell}+\frac{m+n}{\ell}
\quad (m\geq 0 \,,\, n\geq 0)\,,
\end{align}
for $\theta_0=\pi/2$ and for $S^3$\,.

On the other hand, the unpaired bosonic eigenmodes,
$\Delta_\phi \phi =M^2\phi$,
should satisfy
\begin{align}
 \Bigl(\nu-\omega+\frac{1}{\ell}\Bigr)\,\epsilon\phi
-i\gamma^\mu \epsilon\, \da_\mu\phi =0 
\quad\quad (\nu(\nu-2\omega)=M^2) \,,
\end{align} 
which are equivalent to
\begin{align}
 v_+^\mu\da_\mu\phi &=0\,,
\label{3d_chiral_unpaired_scalar}\\
 i v^\mu\da_\mu\phi &=\Bigl(\nu-\omega+\frac{1}{\ell}\Bigr)\phi\,.
\end{align}
It can be easily checked that
these equations imply $\Delta_\phi \phi =M^2\phi$\,.

For $\theta_0<\pi/2$\,,
we can show that there is no nontrivial solution which satisfies 
equation \eqref{3d_chiral_unpaired_scalar} 
and the boundary condition $\phi|=0$\,.
For $\theta_0=\pi/2$\,,
we obtain the following solutions
\begin{align}
 \phi &\propto\cos^{-m}\theta\,\sin^{-n}\theta\,
 e^{i m\varphi-i n\chi}\,,\\
 \nu &=i\sigma_0-\frac{q}{\ell}+\frac{m+n}{\ell}
\quad (m\leq -1 \,,\, n\leq 0)\,.
\end{align}
For $S^3$\,,
we obtain
\begin{align}
 \phi &\propto\cos^{-m}\theta\,\sin^{-n}\theta\,
 e^{i m\varphi-i n\chi}\,,\\
 \nu &=i\sigma_0-\frac{q}{\ell}+\frac{m+n}{\ell}
\quad (m\leq 0 \,,\, n\leq 0)\,.
\end{align}

Therefore, 
up to an overall constant,
the 1-loop determinants for the chiral multiplet are given by\footnote{
$\rho$ are the weights of the representation of the chiral multiplet.}
\begin{align}
 Z^{\rm{1-loop}}_{\rm{chiral}} &=
 \prod_{\rho}\prod_{m}\prod_{n\geq 0}\, (i\rho(\sigma_0)\,\ell -q+2+m-\rho(a_\varphi) +n)\,,
\end{align}
in the case where $\theta_0<\pi/2$\,,
and
\begin{align}
 Z^{\rm{1-loop}}_{\rm{chiral}} &=\prod_{\rho}
 \frac{\prod_{m\geq 0\,, n\geq 0}\, (i\rho(\sigma_0)\,\ell -q+2+m+n)}
{\prod_{m\geq 1\,, n\geq 0}\, (-i\rho(\sigma_0)\,\ell +q+m+n)}\,,
\end{align}
in the case where $\theta_0=\pi/2$\,,
and
\begin{align}
Z^{\rm{1-loop}}_{\rm{chiral}} &=\prod_{\rho}
 \frac{\prod_{m\geq 0\,, n\geq 0}\, (i\rho(\sigma_0)\,\ell -q+2+m+n)}
{\prod_{m\geq 0\,, n\geq 0}\, (-i\rho(\sigma_0)\,\ell +q+m+n)}\,,
\end{align}
in the case for $S^3$\,.
The result for $S^3$ is same as the one in 
\cite{Kapustin:2009kz, Hama:2011ea}.

\subsection{Partition functions and Wilson loops}

Combining the results obtained in this section,
we find that the exact partition function 
for $\theta_0 <\pi/2$ (or $\theta_0 =\pi/2$)
is given by
\begin{eqnarray}
 Z= Z_{\mathrm{classical}} \, 
Z^{\rm{1-loop}}_{\rm{vector}} \, Z^{\rm{1-loop}}_{\rm{chiral}}, 
\end{eqnarray}
where 
\begin{eqnarray}
 Z_{\mathrm{classical}} =e^{ \left(
-i \frac{k}{2 \pi \ell} \tr (\sigma_0)^2
-\frac{2i \zeta}{ \pi \ell^2} \tr \sigma_0 \right) V(\theta_0)},
\end{eqnarray}
where $V(\theta_0)=2 \pi^2 \sin^2 \theta_0 \, \ell^3$.
Because $\sigma_0$ was fixed at the boundary,
there is no matrix integral unlike the case without a boundary.

The supersymmetric Wilson loop operator is given by the following form
\begin{align}
 W_R = \frac{1}{\rm{dim}\,R}
 \tr_R \,\mathrm{P} \exp
 \biggl( \oint_C \rmd\tau 
(i A_\mu \dot x^\mu +\sigma |\dot x|)\biggr)\,,
\end{align}
where $R$ is a representation of the gauge group,
and $\mathrm{P}$ represents path-ordering,
and $C$ denotes a closed world-line parameterized by $x^\mu(\tau)$.
The supersymmetry variation of this operator is
\begin{align}
 \delta W_R \propto \frac{1}{2}
(\bar\epsilon\gamma_\mu\lambda-\bar\lambda\gamma_\mu\epsilon)\dot x^\mu
+\frac{1}{2}
(\bar\epsilon\lambda- \bar\lambda\epsilon) |\dot x|\,.
\end{align}
This vanishes for the Killing spinors \eqref{bks3}
if we take
\begin{align}
 \dot x^\mu = \frac{1}{\ell}(1,-1,0)=-v^\mu\,.
\end{align}
Thus, the supersymmetric Wilson loop is parameterized by
$\theta=\theta_1$ (and $\varphi+\chi=const.$),
and 
we find that the expectation value for this operator is
\begin{eqnarray}
 \langle W_R \rangle = \frac{1}{\rm{dim}\,R}
\tr_R \exp\biggl( 2 \pi 
(i a_\varphi  +\ell\sigma_0)\biggr)\,.
\end{eqnarray}

%%%%%%%%%%%%%%%%%%%%%%%%%%%%%%%%%%%%%%%%%%%%%%%%%%%%%%%%%%%%%
%%%%%%%%%%%%%%%%%%%%%%%%%%%%%%%%%%%%%%%%%%%%%%%%%%%%%%%%%%%%%
%%%%%%%%%%%%%%%%%%%%%%%%%%%%%%%%%%%%%%%%%%%%%%%%%%%%%%%%%%%%%
\section{Two-dimensional theories}
\label{2D-th}

\subsection{A 2D manifold with a boundary}

We will describe
a two dimensional manifold with a boundary
on which the supersymmetric field theories constructed. 
As in the 3D case, we first consider $S^2$.
The coordinates we will use are
$(\theta ,\varphi)$ ($0\leq\theta\leq  \pi ,\,0\leq\varphi<2\pi$ )
with the metric
\begin{align}
\rmd s^2 &= \ell^2(\rmd \theta^2+\sin^2\theta \rmd \varphi^2)\,,\\
\sqrt{g}&=\ell^2\sin\theta\,.
\end{align}
We choose the following zweibein
\begin{align}
e^1=\ell\rmd\theta\,,\quad\,e^2=\ell\sin\theta\,\rmd\varphi \,.
\end{align}
The spin connection is given by
\begin{align}
\omega^{ab}=-\varepsilon^{ab}\cos\theta\,\rmd\varphi\,,\quad(\varepsilon^{12}=1)\,,
\end{align}
and the gamma matrices are
\begin{align}
\gamma^\theta=\frac{1}{\ell}\gamma^1\,,\quad
\gamma^\varphi=\frac{1}{\ell\sin\theta}\gamma^2\,. 
\end{align}

The manifold with the boundary is defined by
just restricting the coordinate $\theta$ as
\begin{eqnarray}
 0\leq\theta\leq\theta_0\,,
\end{eqnarray}
where
$0<\theta_0\leq\pi$.
Thus the boundary defined by $\theta=\theta_0$ is 
a circle parameterized by $\varphi$, except for $\theta_0=\pi$.

\subsection{2D supersymmetric field theories}
Now we will construct the supersymmetric field theories
on the 2-dimensional manifold with the boundary.

First, we will summarize the supersymmetry transformations and 
the supersymmetric invariant Lagrangians of 
the ${\cal N}=(2,2)$ supersymmetric filed theories
on the round $S^2$ \cite{Benini:2012ui,Doroud:2012xw}.

The (positive) Killing spinors on $S^2$ is given by
\begin{align}
D_\mu \epsilon = \frac{i}{2\ell}\gamma_\mu \epsilon\,,
\end{align}
which is solved with constants $C_1, C_2$ as
\begin{align}
\epsilon = 
C_1 e^{i\frac{\varphi}{2}}
\begin{pmatrix}
i\cos\frac{\theta}{2}\\
-\sin\frac{\theta}{2}
\end{pmatrix}
+C_2 e^{-i\frac{\varphi}{2}}
\begin{pmatrix}
-\sin\frac{\theta}{2}\\
i\cos\frac{\theta}{2},
\end{pmatrix}
\end{align}
in our basis.

The supersymmetry transformations of the vector multiplets 
with the Grassmann odd Killing spinor parameters $\epsilon, \bar{\epsilon}$
are the followings:
\begin{align}
\begin{split}
\label{2d-vector_susy_trsf}
\delta A_{\mu} 
&=-\frac{i}{2}(\bar{\epsilon}\gamma_{\mu}\lambda-\bar{\lambda}\gamma_{\mu}\epsilon)\,,\quad
\delta\sigma_1 
=\frac{1}{2}(\bar{\epsilon}\lambda-\bar{\lambda}\epsilon)\,,\quad
\delta\sigma_2 
=-\frac{i}{2}(\bar{\epsilon}\gamma_{3}\lambda-\bar{\lambda}\gamma_{3}\epsilon)
\,,\\
\delta\lambda 
&=(i\gamma_3 F_{12}-D+i\gamma^{\mu}D_{\mu}\sigma_1 
-\gamma_3\gamma^{\mu}D_{\mu}\sigma_2
-\gamma_3[\sigma_1,\sigma_2])\epsilon
+i\sigma_1\gamma^{\mu}D_{\mu}\epsilon-\sigma_2\gamma_3\gamma^{\mu}D_{\mu}\epsilon
\,,\\
\delta\bar{\lambda} &=
(i\gamma_3 F_{12}+D-i\gamma^{\mu}D_{\mu}\sigma_1 -\gamma_3\gamma^{\mu}D_{\mu}\sigma_2
+\gamma_3[\sigma_1,\sigma_2])\bar{\epsilon}
-i\sigma_1\gamma^{\mu}D_{\mu}\bar{\epsilon}
-\sigma_2\gamma_3\gamma^{\mu}D_{\mu}\bar{\epsilon}
\,,\\
\delta D &=
-\frac{i}{2}\bar{\epsilon}(\gamma^{\mu}D_{\mu}\lambda-[\lambda,\sigma_1]
+i[\gamma_3\lambda,\sigma_2])
-\frac{i}{2}(D_{\mu}\bar{\lambda}\gamma^{\mu}-[\bar{\lambda},\sigma_1]
+i[\bar{\lambda}\gamma_3,\sigma_2])\epsilon \\
&\quad -\frac{i}{2}(D_{\mu}\bar{\epsilon}\gamma^{\mu}\lambda
+\bar{\lambda}\gamma^{\mu}D_{\mu}\epsilon)\,.
\end{split}
\end{align}
For a chiral multiplet of R-charge $q$, the supersymmetry
 transformations
are:
\begin{align}
\begin{split}
\label{2d-chiral_susy_trsf}
\delta \phi &= \bar{\epsilon} \psi \,,\qquad
\delta \psi 
= \Bigl( i D_{\mu} \phi\gamma^{\mu} + i \sigma_1\phi + 
\gamma_3 \sigma_2\phi \Bigr) \epsilon 
+\frac{i q}{2}\phi\gamma^{\mu}D_{\mu}\epsilon+ \bar{\epsilon} F \,,\\
\delta \bar{\phi} &= \epsilon\bar{\psi} \,,\qquad
\delta \bar{\psi} 
= \Bigl( i D_{\mu}\bar{\phi}\gamma^{\mu} 
+ i \bar\phi\sigma_1 - \gamma_3 \bar{\phi}\sigma_2  \Bigr) \bar{\epsilon} 
+\frac{i q}{2}\bar{\phi}\gamma^{\mu}D_{\mu}\bar{\epsilon}+ \epsilon \bar{F} \,,\\
\delta F &=
\epsilon \Bigl( i \gamma^{\mu} D_{\mu} \psi - i \sigma_1 \psi 
+ \gamma_3 \sigma_2 \psi - i \lambda \phi \Bigr) 
+\frac{i q}{2}D_{\mu}\epsilon \gamma^{\mu}\psi\,,\\
\delta \bar{F} &=
\bar{\epsilon} \Bigl( i \gamma^{\mu}D_{\mu}\bar{\psi} -i \bar{\psi} \sigma_1 
- \gamma_3 \bar\psi \sigma_2 + i \bar{\phi} \bar{\lambda} \Bigr)
+\frac{i q}{2}D_{\mu}\bar{\epsilon} \gamma^{\mu}\bar{\psi} \;. 
\end{split}
\end{align}

We can construct several actions which are
invariant under the SUSY transformations 
\eqref{2d-vector_susy_trsf}-\eqref{2d-chiral_susy_trsf}.
The first is the Yang-Mills Lagrangian:\footnote{Note that 
the fermion kinetic terms are taken to be symmetrical with respect to 
$\lambda$ and $\bar\lambda$. } 
\begin{align}
\frac{1}{\gyms}\mathcal{L}_{\mathrm{YM}}
&=\frac{1}{\gyms}\tr\Bigl(
\frac{1}{2}(F_{12}-\sigma_2/\ell)^2+\frac{1}{2}D_{\mu}\sigma_1 D^{\mu}\sigma_1
+\frac{1}{2}D_{\mu}\sigma_2 D^{\mu}\sigma_2+\frac{1}{2}(D+\sigma_1 /\ell)^2
-\frac{1}{2}[\sigma_1,\sigma_2]^2\nn\\
&\qquad\qquad\quad +\frac{i}{4}\bar{\lambda}\gamma^{\mu}D_{\mu}\lambda
+\frac{i}{4}\lambda\gamma^{\mu}D_{\mu}\bar{\lambda}
+\frac{i}{2}\bar{\lambda}[\sigma_1,\lambda]
+\frac{1}{2}\bar{\lambda}\gamma^3[\sigma_2,\lambda]
\Bigr)\,,
\end{align}
where $g_{\mathrm{YM}}$ is the coupling constant, 
and $F_{12}$ means 
$\frac{1}{2}\varepsilon^{ab}F_{ab}$.
We can also consider
the Fayet Iliopoulos (FI) term:
\begin{align}
\mathcal{L}_{\mathrm{FI}}&=
\tr\Bigl(-i\zeta D +\frac{i \Theta}{2\pi}F_{12}\Bigr)\,.
\end{align} 
The kinetic terms for a chiral multiplet of R-charge $q$ are
\begin{align}
\mathcal{L}_{\mathrm{mat}}
&=D_{\mu}\bar{\phi}D^{\mu}\phi+\bar{\phi}\sigma_1^2\phi+\bar{\phi}\sigma_2^2\phi
+\frac{iq}{\ell}\bar{\phi}\sigma_1\phi+\frac{q(2-q)}{4\ell^2}\bar{\phi}\phi
+i\bar{\phi}D\phi+\bar{F}F\nn\\
&\quad-\frac{i}{2}\bar{\psi}\gamma^{\mu}D_{\mu}\psi
+\frac{i}{2}D_{\mu}\bar{\psi}\gamma^{\mu}\psi
+i\bar{\psi}\sigma_1\psi-\bar{\psi}\gamma^3\sigma_2\psi
-\frac{q}{2\ell}\bar{\psi}{\psi}
+i\bar{\psi}\lambda\phi -i\bar{\phi}\bar{\lambda}\psi\,.
\end{align}

\subsubsection{The boundary condition}

We study only Dirichlet boundary conditions similar to the 3D theories.

%Our ansatz for 
The boundary conditions we will impose for the vector multiplet
are
\begin{align}
A_{\varphi}|_{\theta=\theta_0} &=a_{\varphi}\,,
\label{2d-bc-a}
\\
\sigma_1 |_{\theta=\theta_0} &= \sigma_0\,,\\
\sigma_2 |_{\theta=\theta_0} &= \eta_0\,,\\
-\ell e^{-i\varphi}\gamma^\theta\lambda|_{\theta=\theta_0} &=\bar\lambda|_{\theta=\theta_0}
\label{2d-bc-la} \,,
\end{align}
where $a_\varphi$, $\sigma_0$ and $\eta_0$ are constants and  commute with each other,
and we consider that they are in the Cartan part of the adjoint representation.
We do not impose any condition for the other fields ( $A_\theta$ and $D$ ).
We take the following boundary conditions for chiral multiplets:
\begin{align}
\begin{split}
\label{2d_chiral_bc}
\phi |_{\theta=\theta_0}&=0\,,\quad
e^{i\frac{\theta_0}{2}\gamma^1}\gamma^3 e^{-i\frac{\theta_0}{2}\gamma^1}\psi |_{\theta=\theta_0}
=-\psi |_{\theta=\theta_0} \,,
\\
\bar\phi |_{\theta=\theta_0}&=0\,,\quad
e^{i\frac{\theta_0}{2}\gamma^1}\gamma^3 e^{-i\frac{\theta_0}{2}\gamma^1}\bar\psi |_{\theta=\theta_0}
  =\bar\psi |_{\theta=\theta_0} \,. 
\end{split}
\end{align}

Under these boundary conditions, 
the half (or $1/4$) of the SUSY
is preserved.\footnote{
Only the first equation in (\ref{2d_e_be}) is needed for the 
vector multiples and only the last two equations in 
(\ref{2d_e_be}) are needed for the chiral multiplets.
If there are only vector multiplets or only chiral multiplets,
the half of the SUSY preserved, however, if both vector multiplets
and chiral multiplets, the $1/4$ of the SUSY
is preserved.}
Indeed,
we can see that
the positive Killing spinors 
which satisfy the relations
\begin{align}
\ell e^{-i\varphi}\gamma^\theta\epsilon =\bar\epsilon \,,\quad
e^{i\frac{\theta}{2}\gamma^1}\gamma^3 e^{-i\frac{\theta}{2}\gamma^1}\epsilon=\epsilon\,,\quad
e^{i\frac{\theta}{2}\gamma^1}\gamma^3 e^{-i\frac{\theta}{2}\gamma^1}\bar\epsilon=-\bar\epsilon\,,
\label{2d_e_be}
\end{align}
generate the supersymmetry transformation 
which is consistent with the above boundary conditions
and under which the actions are invariant (see Appendix
\ref{susy-vari}).

The Grassmann even Killing spinors satisfying the relations \eqref{2d_e_be}
are given by
\begin{align}
\epsilon =e^{i\frac{\varphi}{2}}
\begin{pmatrix}
i\cos\frac{\theta}{2}\\
-\sin\frac{\theta}{2}
\end{pmatrix}  \qquad \text{and} \qquad
\bar{\epsilon} =e^{-i\frac{\varphi}{2}}
\begin{pmatrix}
-\sin\frac{\theta}{2}\\
i\cos\frac{\theta}{2}
\end{pmatrix}\,.\label{bks2}
\end{align}
We can compute the following bi-linears of the spinors:
\begin{align}
\bar\epsilon\epsilon &= 1\,,\\
\bar\epsilon\gamma^3\epsilon &=\cos\theta \equiv w \,,\\
\bar\epsilon\gamma^a\epsilon &=(0,\sin\theta) \equiv v^a \,,
\end{align}
which will be used later.

We should check also that the boundary conditions are consistent with the variational principle.
The surface terms from variation of the Yang-Mills action are
\begin{align}
\frac{1}{\gyms}\int_{\theta=\theta_0}\rmd\varphi\sin\theta\,\tr\Bigl(
\delta A_\varphi \frac{F^{\theta\varphi}}{\sin\theta}-\delta A^{\varphi}\sigma_2
+\delta\sigma_1 D_{\theta}\sigma_1+\delta\sigma_2 D_{\theta}\sigma_2
+\frac{i}{4}\bar{\lambda}\gamma_{\theta}\delta\lambda
+\frac{i}{4}\lambda\gamma_{\theta}\delta\bar{\lambda} 
 \Bigr) \,.
\end{align}
which actually vanish for the above boundary conditions.
The surface term for the FI term is
\begin{align}
 \int_{\theta=\theta_0}\rmd\varphi \,\frac{i \Theta}{2\pi}\, \tr\,\delta
 A_{\varphi},
\end{align}
and the ones for the matter kinetic terms are
\begin{align}
\int_{\theta=\theta_0}\rmd\varphi
\sin\theta\Bigl(\delta\bar\phi\,D_\theta\phi+D_\theta\bar\phi\,\delta\phi
-\frac{i}{2}\bar\psi\gamma_\theta\delta\psi
+\frac{i}{2}\delta\bar\psi\gamma_\theta\psi\Bigr) \,.
\end{align}
We can see that these surface terms vanish with the boundary conditions.

%%%%%%%%%%%%%%%%%%%%%%%%%%%%%%%%%%%%%%%%%%%%%%%%%%%%%%%%%%%%%
\subsection{Localization}

In this subsection, as in the 3D theories, we will construct the 
$\delta$-exact term.

\subsubsection{2D vector multiplet}

%$\epsilon,\bar\epsilon$ are positive Killing spinors which we assume to
%be Grassmann odd.

For the vector multiplet, 
we consider the following $\delta$-exact term 
(ignoring the trace for notational convenience)
\begin{align}
\delta V_{\mathrm{vector}} = \frac{1}{4}\,\delta((\delta'\lambda)^{\dagger}\lambda
 +\bar\lambda(\delta'\bar\lambda)^{\dagger})\,,
\end{align}
where
\begin{align}
 (\delta\lambda)^{\dagger}&=\bar\epsilon\Bigl(
 -i \gamma^3 F_{12}-D-i \gamma^{\mu}D_{\mu}\sigma_1
 +\gamma^3 \gamma^{\mu}D_{\mu}\sigma_2 +\gamma^3[\sigma_1,\sigma_2]
 -\frac{1}{\ell}\sigma_1+\frac{i}{\ell}\gamma^3\sigma_2
 \Bigr)\,,\\
 (\delta\bar\lambda)^{\dagger}&=-\Bigl(
 i \gamma^3 F_{12}+D-i \gamma^{\mu}D_{\mu}\sigma_1
 -\gamma^3 \gamma^{\mu}D_{\mu}\sigma_2 +\gamma^3[\sigma_1,\sigma_2]
 +\frac{1}{\ell}\sigma_1-\frac{i}{\ell}\gamma^3\sigma_2
 \Bigr)\epsilon\,.
\end{align}
Then, 
the bosonic part of $\delta$-exact term is computed using
\begin{align}
 (\delta'\lambda)^{\dagger}\delta\lambda&=\bar\epsilon'\epsilon\Bigl[
 (F_{12}-\sigma_2/\ell)^2 +D_{\mu}\sigma_1 D^{\mu}\sigma_1
+D_{\mu}\sigma_2 D^{\mu}\sigma_2 +(D+\sigma_1 /\ell)^2
-[\sigma_1,\sigma_2]^2 
\nn\\
&\qquad\quad +2 \varepsilon^{\mu\nu}D_\mu\sigma_1 D_\nu\sigma_2
+2 i F_{12}[\sigma_1,\sigma_2]-\frac{2 i}{\ell}[\sigma_1,\sigma_2]\sigma_2
 \Bigr]\,,
\\
 \delta\bar\lambda(\delta'\bar\lambda)^{\dagger}&=\bar\epsilon'\epsilon\Bigl[
 (F_{12}-\sigma_2/\ell)^2 +D_{\mu}\sigma_1 D^{\mu}\sigma_1
+D_{\mu}\sigma_2 D^{\mu}\sigma_2 +(D+\sigma_1 /\ell)^2
-[\sigma_1,\sigma_2]^2 
\nn\\
&\qquad\quad -2 \varepsilon^{\mu\nu}D_\mu\sigma_1 D_\nu\sigma_2
-2 i F_{12}[\sigma_1,\sigma_2]+\frac{2 i}{\ell}[\sigma_1,\sigma_2]\sigma_2
 \Bigr]\,,
\end{align}
where we have used the fact that
$\bar\epsilon\epsilon'=-\bar\epsilon'\epsilon$ by \eqref{2d_e_be},
as
\begin{align}
  &(\delta'\lambda)^{\dagger}\delta\lambda+\delta\bar\lambda(\delta'\bar\lambda)^{\dagger}\nn\\
  &=2\bar\epsilon'\epsilon\Bigl[
 (F_{12}-\sigma_2/\ell)^2 +D_{\mu}\sigma_1 D^{\mu}\sigma_1
+D_{\mu}\sigma_2 D^{\mu}\sigma_2 +(D+\sigma_1 /\ell)^2
-[\sigma_1,\sigma_2]^2 
 \Bigr]
\\
  &=4\bar\epsilon'\epsilon \,\mathcal{L}_{\mathrm{YM}}^{\mathrm{boson}}\,.
\end{align}

Next we will compute the fermionic part of the $\delta$-exact term.
In
\begin{align}
 \delta((\delta'\lambda)^{\dagger})\lambda= \bar\epsilon'\epsilon\Bigl(
 i\lambda\gamma^\mu D_\mu\bar\lambda-i\lambda[\sigma_1,\bar\lambda]
 +\lambda\gamma^3[\sigma_2,\bar\lambda]
 \Bigr)+(\bar\epsilon'\gamma^\mu\bar\epsilon)(i(D_\mu\lambda)\lambda)\,,
\end{align}
the last term is total derivative and it is equivalent to 
the surface term:
\begin{align}
 i(\bar\epsilon'\gamma^\theta\bar\epsilon)(\lambda\lambda)|
 =i\bar\epsilon'\epsilon(\bar\lambda\gamma^\theta\lambda)|\,,
\end{align}
where we have used \eqref{2d-bc-la} and \eqref{2d_e_be}.
Similarly, in
\begin{align}
 \bar\lambda\delta((\delta'\bar\lambda)^{\dagger})&=
 \bar\epsilon'\epsilon\Bigl(
 i\bar\lambda\gamma^\mu D_\mu\lambda
 +i\bar\lambda[\sigma_1,\lambda]
 +\bar\lambda\gamma^3[\sigma_2,\lambda]
 \Bigr)
 -(\epsilon'\gamma^\mu\epsilon)(i(D_\mu\bar\lambda)\bar\lambda)\,,
\end{align}
the last term is also a total derivative and 
it becomes the surface term:
\begin{align}
 i(\epsilon'\gamma^\theta\epsilon)(\bar\lambda\bar\lambda)|
 =i\bar\epsilon'\epsilon(\bar\lambda\gamma^\theta\lambda)|\,,
\end{align}
where we have used \eqref{2d-bc-la} and \eqref{2d_e_be}.
Thus, the fermionic part of $\delta$-exact term is
\begin{align}
 \delta((\delta'\lambda)^{\dagger})\lambda
 +\bar\lambda\delta((\delta'\bar\lambda)^{\dagger})
 &=\bar\epsilon'\epsilon\Bigl(
 i\bar\lambda\gamma^\mu D_\mu\lambda
 +i\lambda\gamma^\mu D_\mu\bar\lambda
 +2 i\bar\lambda[\sigma_1,\lambda]
 +2\bar\lambda\gamma^3[\sigma_2,\lambda]
 \Bigr)\\
 &=4\bar\epsilon'\epsilon\,\mathcal{L}_{\mathrm{YM}}^{\mathrm{fermion}}\,,
\end{align}
where no surface terms present.
Therefore, we find that $\delta V_{\mathrm{vector}} = {\cal L}_{YM}$.

The saddle point of the bosonic part of this $\delta V_{\mathrm{vector}}$ 
is given by
\begin{align}
F_{12}=\frac{\sigma_2}{\ell}\,,\,\, D_\mu \sigma_1=D_\mu \sigma_2=0\,,
\,\,D=-\frac{\sigma_1}{\ell}\,,\,\,[\sigma_1,\sigma_2]=0\,.
\end{align}
Using the boundary condition and choosing a gauge condition,
the solutions of these equations 
are
\begin{align}
A=\ell\eta_0(\kappa-\cos\theta)\rmd\varphi=a_\varphi\rmd\varphi\,,
\quad
\sigma_1=-\ell D=\sigma_0\,,
\quad
\sigma_2=\eta_0\,,
\end{align}
where $\kappa=1$ $(\kappa=-1)$ for the patch covering  
$S^2$ except a point $\theta=\pi$ $(\theta=0)$.\footnote{
We can take $\kappa=1$ for the case where $\theta_0<\pi$.}
In the case for $S^2$ \cite{Benini:2012ui, Doroud:2012xw}, 
since the flux $2\ell^2 F_{12}$ is GNO quantized \cite{Goddard:1976qe},
$\rho(2\ell\eta_0)$ should be an integer
for any representation $R$ of the gauge group $G$
and any weight $\rho\in R$.
For $\theta_0<\pi$, however,
the flux does not need to be quantized.

\subsubsection{2D chiral multiplet}

For the chiral multiplet, 
we consider the following $\delta$-exact term
\begin{align}
\delta V_{\mathrm{chiral}}&=\frac{1}{2}\, 
\delta[(\delta'\psi)^\dagger\psi+\bar\psi(\delta'\bar\psi)^\dagger] 
+\frac{q-1}{2 \ell}\,\delta[\bar\phi\,\delta'\phi-(\delta'\bar\phi)\phi]\,, 
\end{align}
where
\begin{align}
(\delta'\psi)^\dagger &\equiv\bar\epsilon'\Bigl(-i D_\mu\bar\phi\gamma^\mu
-i \bar\phi\,\sigma_1 +\bar\phi\sigma_2\gamma^3-\frac{q}{2\ell}\bar\phi\Bigr)-\epsilon' \bar F\,,
\\
(\delta'\bar\psi)^\dagger &\equiv-\Bigl(i D_\mu\phi\gamma^\mu
-i\sigma_1\phi +\sigma_2\phi\gamma^3-\frac{q}{2\ell}\phi\Bigr)\epsilon'+\bar\epsilon' F\,.
\end{align}

The bosonic part of $\delta V_{\mathrm{chiral}}$ is given by
\begin{align}
 &\frac{1}{2}\,(\delta'\psi)^\dagger\delta\psi +\delta\bar\psi(\delta'\bar\psi)^\dagger
 +\frac{q-1}{2\ell}\,[\bar\phi\,(\delta\,\delta'\phi)-(\delta\,\delta'\bar\phi)\phi]
 \nn\\
 &=\bar\epsilon'\epsilon\Bigl(
 D_\mu\bar\phi D^\mu\phi+\bar\phi\,\sigma_1^2\phi +\bar\phi\,\sigma_2^2\phi
+i\frac{q-1}{\ell}\bar\phi\,\sigma_1\phi
-\frac{q(q-2)}{4\ell^2}\,\bar\phi\,\phi+\bar F F
\Bigr)
\nn\\
&\quad + i \,\bar\epsilon'\gamma^\mu\epsilon\Bigl[
\frac{1}{2\ell}(D_\mu\bar\phi\,\phi -\bar\phi D_\mu\phi)
 -i\varepsilon_{\mu\nu}(D^\nu\bar\phi\,\sigma_2\phi+\bar\phi\,\sigma_2 D^\nu\phi)
 \Bigr]
 \nn\\
  &\quad + i\,\bar\epsilon'\gamma^3\epsilon\Bigl(
i\varepsilon^{\mu\nu}D_\mu\bar\phi\,D_\nu\phi
+\frac{i}{\ell}\bar\phi\,\sigma_2\phi
\Bigr)\,.
\end{align} 
Similarly, the fermionic part can be computed as
\begin{align}
&\frac{1}{2}\delta((\delta'\psi)^\dagger)\psi +\bar\psi\delta((\delta'\bar\psi)^\dagger)
+\frac{q-1}{2\ell}\,[\delta\bar\phi\,\delta'\phi -\delta'\bar\phi\,\delta\phi]
\nn\\
&=\bar\epsilon'\epsilon\Bigl(
  \frac{i}{2} D_\mu\bar\psi\gamma^\mu\psi -\frac{i}{2} \bar\psi\gamma^\mu D_\mu\psi
+ i\bar\psi\,\sigma_1\psi -\bar\psi\,\sigma_2\gamma^3\psi
-\frac{3 i}{4}\bar\phi\bar\lambda\psi +\frac{3 i}{4}\bar\psi\lambda\phi
-\frac{q}{2\ell}\bar\psi\psi
 \Bigr)
\nn\\
&\quad + i \,\bar\epsilon'\gamma^\mu\epsilon\Bigl[
\frac{i}{2} \varepsilon_{\mu\nu}(D^\nu\bar\psi\,\gamma^3\psi +\bar\psi\,\gamma^3 D^\nu\psi)
+\frac{1}{4}\bar\phi\bar\lambda\gamma_\mu\psi -\frac{1}{4}\bar\psi\gamma_\mu\lambda\phi
-\frac{i}{2\ell}\bar\psi\gamma_\mu\psi
 \Bigr]
 \nn\\
 &\quad + i\,\bar\epsilon'\gamma^3\epsilon\Bigl(
 \frac{1}{2} D_\mu\bar\psi\,\gamma_3\gamma^\mu\psi +\frac{1}{2}\bar\psi\,\gamma_3\gamma^\mu D_\mu\psi
+\frac{1}{4}\bar\phi\bar\lambda\gamma_3\psi -\frac{1}{4}\bar\psi\gamma_3\lambda\phi
-\frac{i}{\ell}\bar\psi\gamma_3\psi
 \Bigr)
 \nn\\
 &\quad
-\frac{i}{4}\epsilon'\gamma^\mu\epsilon\,
\bar\psi\gamma_\mu\bar\lambda\phi
+\frac{i}{4}\bar\epsilon'\gamma^\mu\bar\epsilon\,
\bar\phi\lambda\gamma_\mu\psi
-\frac{i}{4}\epsilon'\gamma^3\epsilon\,
\bar\psi\gamma_3\bar\lambda\phi
+\frac{i}{4}\bar\epsilon'\gamma^3\bar\epsilon\,
\bar\phi\lambda\gamma_3\psi\,.
\end{align} 
Therefore,
\begin{align}
&\delta V_{\mathrm{chiral}}
= \bar\epsilon'\epsilon\Bigl(
 D_\mu\bar\phi D^\mu\phi+\bar\phi\,\sigma_1^2\phi +\bar\phi\,\sigma_2^2\phi
+i\frac{q-1}{\ell}\bar\phi\,\sigma_1\phi
-\frac{q(q-2)}{4\ell^2}\,\bar\phi\,\phi+\bar F F
\nn\\ 
&\qquad\qquad+\frac{i}{2} D_\mu\bar\psi\gamma^\mu\psi -\frac{i}{2} \bar\psi\gamma^\mu D_\mu\psi
+ i\bar\psi\,\sigma_1\psi -\bar\psi\,\sigma_2\gamma^3\psi
-\frac{3 i}{4}\bar\phi\bar\lambda\psi +\frac{3 i}{4}\bar\psi\lambda\phi
-\frac{q}{2\ell}\bar\psi\psi
 \Bigr)
\nn\\
&\quad + i \,\bar\epsilon'\gamma^\mu\epsilon\Bigl[
\frac{1}{2\ell}(D_\mu\bar\phi\,\phi -\bar\phi D_\mu\phi)
 -i\varepsilon_{\mu\nu}(D^\nu\bar\phi\,\sigma_2\phi+\bar\phi\,\sigma_2 D^\nu\phi)
\nn\\
&\qquad\qquad\qquad
+\frac{i}{2} \varepsilon_{\mu\nu}(D^\nu\bar\psi\,\gamma^3\psi +\bar\psi\,\gamma^3 D^\nu\psi)
+\frac{1}{4}\bar\phi\bar\lambda\gamma_\mu\psi -\frac{1}{4}\bar\psi\gamma_\mu\lambda\phi
-\frac{i}{2\ell}\bar\psi\gamma_\mu\psi
 \Bigr]
 \nn\\
 &\quad + i\,\bar\epsilon'\gamma^3\epsilon\Bigl(
i\varepsilon^{\mu\nu}D_\mu\bar\phi\,D_\nu\phi
+\frac{i}{\ell}\bar\phi\,\sigma_2\phi
\nn\\
&\qquad\qquad\qquad
+\frac{1}{2} D_\mu\bar\psi\,\gamma_3\gamma^\mu\psi +\frac{1}{2}\bar\psi\,\gamma_3\gamma^\mu D_\mu\psi
+\frac{1}{4}\bar\phi\bar\lambda\gamma_3\psi -\frac{1}{4}\bar\psi\gamma_3\lambda\phi
-\frac{i}{\ell}\bar\psi\gamma_3\psi
 \Bigr)
 \nn\\
 &\quad
-\frac{i}{4}\epsilon'\gamma^\mu\epsilon\,
\bar\psi\gamma_\mu\bar\lambda\phi
+\frac{i}{4}\bar\epsilon'\gamma^\mu\bar\epsilon\,
\bar\phi\lambda\gamma_\mu\psi
-\frac{i}{4}\epsilon'\gamma^3\epsilon\,
\bar\psi\gamma_3\bar\lambda\phi
+\frac{i}{4}\bar\epsilon'\gamma^3\bar\epsilon\,
\bar\phi\lambda\gamma_3\psi\,.
\end{align}
If we use the $\delta$-exact term
$\delta[(\delta'\psi)^\dagger\psi+\bar\psi(\delta'\bar\psi)^\dagger]/2$,
which is manifestly positive definite,
instead of $\delta V_{\mathrm{chiral}}$,
the saddle point is given by $\phi=F=\bar\phi=\bar F=0$\,.
Because the addition of
$\delta[\bar\phi\,\delta'\phi-(\delta'\bar\phi)\phi]\,(q-1)/(2 \ell)$
will not change the 1-loop determinant,
we will use  $\delta V_{\mathrm{chiral}}$ for simplicity of later computation.

\subsection{One-loop determinant}

%In this section, we assume $\epsilon$ and $\bar\epsilon$ as
%Grassmann-even.

In this subsection, 
we will compute the 1-loop determinant for the $\delta$-exact action
in the similar way as in subsection \ref{3d-1loop}.
The computation will be basically follow \cite{Gomis:2012wy}.

\subsubsection{Vector multiplet}
We will compute the 1-loop determinant for the vector multiplet. 

We expand the fields around the saddle point as
\begin{align}
\begin{split}
A_\mu&=a_\mu +\frac{1}{\sqrt{t}}\hat{A}_\mu\,,
\quad
\sigma_1=\sigma_0+ \frac{1}{\sqrt{t}}\hat\sigma_1\,,
\\
\sigma_2&=\eta_0+ \frac{1}{\sqrt{t}}\hat\sigma_2\,,
\quad
D=-\frac{\sigma_0}{\ell}+\frac{1}{\sqrt{t}}\hat{D} 
\\
\lambda&=\frac{1}{\sqrt{t}}\hat\lambda \,,
\quad
\bar\lambda=\frac{1}{\sqrt{t}}\hat{\bar\lambda} \,,
\end{split}
\end{align}
where
\begin{align}
a_\theta=0\,,
\quad
a_\varphi=\ell\,\eta_0(\kappa-\cos\theta) \,.
\end{align}
The boundary conditions for the fluctuation fields at $\theta=\theta_0$ are
\begin{align}
 \hat{A}_\varphi |=\hat\sigma_1 |=\hat\sigma_2 |=0\,.
 \label{2d_hat_bd_cond}
\end{align}
Then, performing the integral by parts,
we have
\begin{align}
t\int\rmd^2 x\sqrt{g} \,\mathcal{L}_{\mathrm{YM}}
&=\int\rmd^2 x\sqrt{g} \,\tr\Bigl(
-\frac{1}{2}A^\mu(\ast \da\ast \da A)_\mu-\varepsilon^{\mu\nu}\frac{\sigma_2}{\ell}\da_\mu A_\nu
\nn\\
&\qquad\qquad\qquad\quad+\frac{1}{2}[\sigma_0,A^\mu][\sigma_0,A_\mu]
+\frac{1}{2}[\eta_0,A^\mu][\eta_0,A_\mu]
\nn\\
&\qquad\qquad\qquad\quad-\frac{1}{2}\sigma_1(\ast \da\ast \da \sigma_1)
+i[\sigma_0,A^\mu]\da_\mu\sigma_1
\nn\\
&\qquad\qquad\qquad\quad-\frac{1}{2}\sigma_2(\ast \da\ast \da \sigma_2)
+i[\eta_0,A^\mu]\da_\mu\sigma_2
+\frac{\sigma_2^2}{2\ell^2}
\nn\\
&\qquad\qquad\qquad\quad
+\frac{1}{2}(D+\sigma_1/\ell)^2
-\frac{1}{2}([\sigma_0,\sigma_2]-[\eta_0,\sigma_1])^2
\nn\\
&\qquad\qquad\qquad\quad
+\frac{i}{4}\bar{\lambda}\gamma^{\mu}\da_{\mu}\lambda
+\frac{i}{4}\lambda\gamma^{\mu}\da_{\mu}\bar{\lambda}
+\frac{i}{2}\bar{\lambda}[\sigma_0,\lambda]
+\frac{1}{2}\bar{\lambda}\gamma^3[\eta_0,\lambda]
\Bigr)
\nn\\
&\qquad\qquad\qquad+\mathcal{O}(t^{-1/2})\,,
\label{2d_Delta_b}
\end{align}
where the covariant derivative $\da$ means $\da_\mu=\nabla_\mu-i[a_\mu,\cdot\,]$\,
and we have omitted the hat symbols for the fluctuation fields.

As explained in \cite{Gomis:2012wy},
for evaluating the 1-loop factor for bosonic fields,
only the eigenvalues of 
the bosonic eigenmodes which are orthogonal 
to the ``non-physical modes'' should be included.
Using the results in \cite{Gomis:2012wy}, we can easily see 
that 
the orthogonal conditions are given by
\begin{align}
\ast\da\ast A =i[\eta_0,\sigma_2]\,,\quad \sigma_1=0\,,
\end{align}
for our conventions.
The latter condition, $\sigma_1=0$, implies that we need to
consider only $A_\mu$ and $\sigma_2$.\footnote{
The eigenvalue problem for the bosonic kinetic operator in \eqref{2d_Delta_b}
 is also consistent with $\sigma_1=0$
if the former condition, $\ast\da\ast A =i[\eta_0,\sigma_2]$, is imposed.}

Using the Cartan decomposition,
all the adjoint fields $X$
can be decomposed as
\begin{align}
X=\sum_i X^i H_i+\sum_{\alpha\in\Delta_+}(X^\alpha E_\alpha + X^{-\alpha} E_{-\alpha})\,,
\end{align}
where $H_i$ are the Cartan generators 
and $\Delta_+$ is the set of the positive roots.
The generators $E_\alpha$ are normalized as
$\tr(E_\alpha E_\beta)=\delta_{\alpha+\beta}$.
Then,
the quadratic terms we consider have the following forms\footnote{
The contributions from the Cartan part $X^i$ are the form $\ell^{p}$,
thus, we have neglected them.}
\begin{align}
 (A^{-\alpha},\sigma_2^{-\alpha})\,\Delta_b
 \begin{pmatrix}
A^{\alpha}\\
\sigma_2^{\alpha}
\end{pmatrix}
+\frac{1}{2}\,(\lambda^{-\alpha},\bar\lambda^{-\alpha})
\begin{pmatrix}
0&\Delta_{\bar\lambda}\\
\Delta_\lambda&0
\end{pmatrix}
\begin{pmatrix}
\lambda^{\alpha}\\
\bar\lambda^{\alpha}
\end{pmatrix}\,,
\end{align}
where
\begin{align}
\Delta_b&=
\begin{pmatrix}
-\ast\da\ast\da+\alpha(\sigma_0)^2+\alpha(\eta_0)^2
&-i\alpha(\eta_0)\da+\frac{1}{\ell}\ast\da\\
-i\alpha(\eta_0)\ast\da\ast-\frac{1}{\ell}\ast\da
&-\ast\da\ast\da+\alpha(\sigma_0)^2+\frac{1}{\ell^2}
\end{pmatrix} 
\\
\Delta_\lambda&=i\gamma^\mu\da_\mu+i\alpha(\sigma_0)+\gamma^3\alpha(\eta_0)\,,
\quad
\Delta_{\bar\lambda}=i\gamma^\mu\da_\mu-i\alpha(\sigma_0)+\gamma^3\alpha(\eta_0)\,.
\end{align}
Accordingly, 
we will consider the eigenvalue problem for the above kinetic terms
as in subsection \ref{3d-1loop}.

First, we will construct a correspondence between
the bosonic eigenmodes and fermionic eigenmodes 
as in subsection \ref{3d-1loop} and as in \cite{Gomis:2012wy}.
Let $(A^{\alpha},\sigma_2^{\alpha})$ be an eigenmode for $\Delta_b$:
\begin{align}
\Delta_b 
\begin{pmatrix}
A^{\alpha}\\
\sigma_2^{\alpha}
\end{pmatrix}
=M^2
\begin{pmatrix}
A^{\alpha}\\
\sigma_2^{\alpha}
\end{pmatrix}\,,
\quad
\ast\da\ast A^{\alpha} =i\alpha(\eta_0)\sigma_2^{\alpha}\,.
\end{align}
Then, if we define
\begin{align}
\lambda_1^{\alpha}&\equiv(\gamma^\mu A^{\alpha}_\mu+\gamma^3 \sigma_2^{\alpha})\epsilon\,,
\label{2d-lambda1}
\\
\lambda_2^{\alpha}&\equiv -(\ast\da A^{\alpha})\gamma^3\epsilon
+(\ast\da\sigma_2^{\alpha})_\mu\gamma^\mu\epsilon
+\frac{1}{\ell}\gamma^3\sigma_2^{\alpha}\epsilon
+\alpha(\eta_0)A_\mu^\alpha\gamma^3\gamma^\mu\epsilon \,,
\\
\bar\lambda_1^{\alpha}&\equiv-(\gamma^\mu A^{\alpha}_\mu+\gamma^3 \sigma_2^{\alpha})\bar\epsilon\,,
\\
\bar\lambda_2^{\alpha}&\equiv -(\ast\da A^{\alpha})\gamma^3\bar\epsilon
+(\ast\da\sigma_2^{\alpha})_\mu\gamma^\mu\bar\epsilon
+\frac{1}{\ell}\gamma^3\sigma_2^{\alpha}\bar\epsilon
+\alpha(\eta_0)A_\mu^\alpha\gamma^3\gamma^\mu\bar\epsilon \,,
\label{2d-blambda2}
\end{align}
%we find
%\begin{align}
%\Delta_\lambda 
%\begin{pmatrix}
%\lambda_1^{\alpha}\\
%\lambda_2^{\alpha}
%\end{pmatrix}
%&=
%\begin{pmatrix}
%\alpha(\sigma_0) &1\\
%M^2 -\alpha(\sigma_0) ^2 & i\alpha(\sigma_0) 
%\end{pmatrix}
%\begin{pmatrix}
%\lambda_1^{\alpha}\\
%\lambda_2^{\alpha}
%\end{pmatrix}\,,
%\\
%\Delta_{\bar\lambda} 
%\begin{pmatrix}
%\bar\lambda_1^{\alpha}\\
%\bar\lambda_2^{\alpha}
%\end{pmatrix}
%&=
%\begin{pmatrix}
%-i\alpha(\sigma_0) &-1\\
%-M^2 +\alpha(\sigma_0) ^2 & -i\alpha(\sigma_0) 
%\end{pmatrix}
%\begin{pmatrix}
%\bar\lambda_1^{\alpha}\\
%\bar\lambda_2^{\alpha}
%\end{pmatrix}\,.
%\end{align}
we can obtain fermionic eigenmodes for $\Delta_\lambda$ and $\Delta_{\bar\lambda}$ from
the bosonic eigenmode,
\begin{align}
 \Delta_\lambda\lambda_\pm^{\alpha}&=\nu_\pm\lambda_\pm^{\alpha} \,,
 \quad
 \Delta_{\bar\lambda}\bar\lambda_\pm^{\alpha}=-\nu_\pm\bar\lambda_\pm^{\alpha} \,,
\end{align}
where
\begin{align}
  \lambda_\pm^{\alpha}&\equiv\pm\sqrt{M^2-\alpha(\sigma_0)^2}\lambda_1^{\alpha}+\lambda_2^{\alpha}\,,
  \quad
  \bar\lambda_\pm^{\alpha}\equiv\pm\sqrt{M^2-\alpha(\sigma_0)^2}\bar\lambda_1^{\alpha}+\bar\lambda_2^{\alpha}\,,
  \\
  \nu_\pm&=i\alpha(\sigma_0)\pm\sqrt{M^2-\alpha(\sigma_0)^2}\,.
\end{align}
One can show that
these fermionic eigenmodes
satisfy the boundary condition
\begin{align}
-\ell e^{-i\varphi}\gamma^\theta\lambda_\pm^{\alpha}|=\bar\lambda_\pm^{\alpha}|\,, 
\end{align}
since $(A,\sigma_2)$ satisfies the boundary condition \eqref{2d_hat_bd_cond}.
On the other hand,
if $\lambda$ and $\bar\lambda$ are fermionic eigenmodes,
\begin{align}
 \Delta_\lambda\lambda^{\alpha}&=\nu\lambda^{\alpha} \,,
 \quad
 \Delta_{\bar\lambda}\bar\lambda^{\alpha}=-\nu\bar\lambda^{\alpha} \,,
\end{align}
which satisfy the boundary condition
\begin{align}
 -\ell e^{-i\varphi}\gamma^\theta\lambda^{\alpha}|=\bar\lambda^{\alpha}|\,, 
\end{align}
then we can obtain an eigenmode for $\Delta_b$,
\begin{align}
 A^{\alpha}_\mu&\equiv
(\nu-i\alpha(\sigma_0))(\bar\epsilon\gamma_\mu\lambda^{\alpha}+\epsilon\gamma_\mu\bar\lambda^{\alpha})
 -i\da_\mu(\bar\epsilon\lambda^{\alpha}-\epsilon\bar\lambda^{\alpha})\,,
 \\
 \sigma_2^{\alpha}&\equiv
(\nu-i\alpha(\sigma_0))(\bar\epsilon\gamma_3\lambda^{\alpha}+\epsilon\gamma_3\bar\lambda^{\alpha})
 -\alpha(\eta_0)(\bar\epsilon\lambda^{\alpha}-\epsilon\bar\lambda^{\alpha})\,.
\end{align}
This mode satisfies the boundary conditions \eqref{2d_hat_bd_cond}
and the condition
\begin{align}
\ast\da\ast A^\alpha=i\alpha(\eta_0) \sigma_2^\alpha\,.
\end{align}
The corresponding eigenvalue is
$(\nu-i\alpha(\sigma_0))^2+\alpha(\sigma_0)^2$
\begin{align}
\Delta_b \begin{pmatrix}
A^{\alpha}\\
\sigma_2^{\alpha}
\end{pmatrix}
=[(\nu-i\alpha(\sigma_0))^2+\alpha(\sigma_0)^2]
\begin{pmatrix}
A^{\alpha}\\
\sigma_2^{\alpha}
\end{pmatrix}
\,.
\end{align}
Due to the correspondence between
the bosonic eigenmodes and fermionic eigenmodes,
the eigenvalues for bosonic modes are almost canceled 
by the ones for fermionic eigenmodes in the evaluation for 1-loop determinant.
Therefore, %as 3D theories,
we will consider only the eigenmodes which are not canceled.

\paragraph{Unpaired eigenmodes}~\\

%Secondly, 
We consider unpaired eigenmodes whose eigenvalues are not canceled.
In Appendix \ref{useful_2d}\,, we summarize
some useful formulas for Killing spinors \eqref{bks2} and their bi-linears, 
which will be used in the following calculation.

We consider unpaired bosonic eigenmodes $(A_\mu,\sigma_2)$,
\begin{align}
\Delta_b 
\begin{pmatrix}
A^{\alpha}\\
\sigma_2^{\alpha}
\end{pmatrix}
&=M^2
\begin{pmatrix}
A^{\alpha}\\
\sigma_2^{\alpha}
\end{pmatrix}\,,
\label{2d_eigen_A_sigma}
\end{align} 
which satisfy 
$\ast\da\ast A^{\alpha} =i\alpha(\eta_0)\sigma_2^{\alpha}$ and
\begin{align}
C\lambda_1 +\lambda_2&=0\,,\quad
C\bar\lambda_1+\bar\lambda_2=0\,,
\label{2d_missing_lambda}
\end{align}
where $C=\sqrt{M^2-\alpha(\sigma_0)^2}$ or $C=-\sqrt{M^2-\alpha(\sigma_0)^2}$
and $\lambda_1, \lambda_2, \bar\lambda_1, \bar\lambda_2$  are given by \eqref{2d-lambda1}-\eqref{2d-blambda2}\,.
Accordingly,
$(A_\mu,\sigma_2)$ should satisfy the following equations:
\begin{align}
\begin{split}
 &C(v^\mu A_\mu+w\sigma_2)=0\,,\quad
  i C\tilde{v}^\mu A_\mu-(\ast\da A)+\frac{1}{\ell}\sigma_2=0\,.
 \\
 &C\sigma_2+i\tilde{v}^\mu(\ast\da\sigma_2)_\mu+\alpha(\eta_0)v^{\mu} A_\mu=0\,,
 \\
 &w(\ast\da A)-v^\mu(\ast\da\sigma_2)_\mu-\frac{w}{\ell}\sigma_2-i\alpha(\eta_0)\tilde{v}^\mu A_\mu=0\,,
 \end{split}
\end{align}
where the definitions of $w,\,v^\mu,\,\tilde{v}^\mu$ are given by \eqref{2d-bilinear}.
If we make the ansatz
\begin{align}
A=f_1(\theta)e^{i m\varphi} e^1+f_2(\theta)e^{i m\varphi}e^2 \,,\quad
\sigma_2=f(\theta)e^{i m\varphi}\,,
\end{align}
where $m$ is an integer,
then we can find that 
\begin{align}
\ell C=-m+\kappa\ell\alpha(\eta_0) 
\end{align}
so that the condition \eqref{2d_missing_lambda} should be consistent with the eigenmode equation \eqref{2d_eigen_A_sigma}.
The remaining conditions imply that $f_1$, $f_2$ and $f$ should satisfy the following equations
\begin{align}
\begin{split}
\sin\theta f_2+\cos\theta f &=0\,, 
\\
\partial_\theta(\sin\theta \,f_1)&=-i(\ell\alpha(\eta_0)\cos\theta-\ell C)\,f_2-i\ell\alpha(\eta_0)\frac{\sin^2\theta}{\cos\theta}f_2\,,
\\
\partial_\theta(\sin\theta \,f_2)&=i(\ell\alpha(\eta_0)\cos\theta-\ell C \cos^2\theta)f_1-\frac{\sin^2\theta}{\cos\theta}f_2\,.
\end{split}
\end{align}
One can find that general solutions of these equations are given by
\begin{align}
f_1(\theta)&=-i C_1\Bigl(\sin\frac{\theta}{2}\Bigr)^{-(\kappa-1)\ell\alpha(\eta_0)}
                    \Bigl(\cos\frac{\theta}{2}\Bigr)^{-(\kappa+1)\ell\alpha(\eta_0)}\sin^{m-1}\theta
                    \nn\\
               &\quad+i C_2\Bigl(\sin\frac{\theta}{2}\Bigr)^{(\kappa-1)\ell\alpha(\eta_0)}
                 \Bigl(\cos\frac{\theta}{2}\Bigr)^{(\kappa+1)\ell\alpha(\eta_0)} \sin^{-m-1}\theta\,,
               \\
f_2(\theta)&=C_1\cos\theta\Bigl(\sin\frac{\theta}{2}\Bigr)^{-(\kappa-1)\ell\alpha(\eta_0)}
                    \Bigl(\cos\frac{\theta}{2}\Bigr)^{-(\kappa+1)\ell\alpha(\eta_0)}\sin^{m-1}\theta
                    \nn\\
               &\quad+ C_2\cos\theta\Bigl(\sin\frac{\theta}{2}\Bigr)^{(\kappa-1)\ell\alpha(\eta_0)}
                 \Bigl(\cos\frac{\theta}{2}\Bigr)^{(\kappa+1)\ell\alpha(\eta_0)} \sin^{-m-1}\theta\,,
               \\
f(\theta)&=-C_1\Bigl(\sin\frac{\theta}{2}\Bigr)^{-(\kappa-1)\ell\alpha(\eta_0)}
                    \Bigl(\cos\frac{\theta}{2}\Bigr)^{-(\kappa+1)\ell\alpha(\eta_0)}\sin^{m}\theta
                    \nn\\
               &\quad-C_2\Bigl(\sin\frac{\theta}{2}\Bigr)^{(\kappa-1)\ell\alpha(\eta_0)}
                 \Bigl(\cos\frac{\theta}{2}\Bigr)^{(\kappa+1)\ell\alpha(\eta_0)} \sin^{-m}\theta\,,                
\end{align}
where $C_1$ and $C_2$ are constants.
The regularity at $\theta=0$
requires $C_1=0$ or $C_2=0$.
Thus,
in the case where $\theta_0<\pi$,
there is no nontrivial solution which satisfies the boundary condition.
For the special case that $\theta_0=\pi$, there are solutions:
\begin{align}
 C_2&=0\,,\quad\nn\\
 \ell C&=-m+\ell\alpha(\eta_0)\,,\quad
\left\{ \begin{array}{ll}
m\geq 2 \ell\alpha(\eta_0)+2\,&\text{for}\quad\ell\alpha(\eta_0)\geq 0\\
m\geq 1\,&\text{for}\quad\ell\alpha(\eta_0)<0, \\
\end{array} \right.
\end{align}
or
\begin{align}
 C_1&=0\,,\quad\nn\\
  \ell C&=-m+\ell\alpha(\eta_0)\,,\quad
\left\{ \begin{array}{ll}
m\leq -1\,&\text{for}\quad\ell\alpha(\eta_0)> 0\\
m\leq 2 \ell\alpha(\eta_0)-2\,&\text{for}\quad\ell\alpha(\eta_0)\leq 0,\\
\end{array} \right.
\end{align}
where we assume that $2\ell\alpha(\eta_0)$ takes an integer value. 
For $S^2$,
there are solutions:
\begin{align}
 C_2&=0\,,\quad\nn\\
 \ell C&=-m+\ell\alpha(\eta_0)\,,\quad
\left\{ \begin{array}{ll}
m\geq 2 \ell\alpha(\eta_0)+1\,&\text{for}\quad\ell\alpha(\eta_0)\geq 0\\
m\geq 1\,&\text{for}\quad\ell\alpha(\eta_0)<0, \\
\end{array} \right.
\end{align}
or
\begin{align}
 C_1&=0\,,\quad\nn\\
  \ell C&=-m+\ell\alpha(\eta_0)\,,\quad
\left\{ \begin{array}{ll}
m\leq -1\,&\text{for}\quad\ell\alpha(\eta_0)> 0\\
m\leq 2 \ell\alpha(\eta_0)-1\,&\text{for}\quad\ell\alpha(\eta_0)\leq 0\,.\\
\end{array} \right.
\end{align}
Thus, shifting $m$ appropriately,
$\ell C$ is given by
\begin{align}
 \ell C=
\left\{\begin{array}{ll}
-m-|\ell\alpha(\eta_0)|\,&(m\geq 1)\\
-m+|\ell\alpha(\eta_0)|\,&(m\leq -1).\\
\end{array} \right.
\end{align}

%%%%%%%%%%%%%%%%%%%%%%%%%%%%%%%%%%%%%%%%%%%%%%%%%%%%%%%%%%%%%
Next,
we consider unpaired fermionic eigenmodes,
\begin{align}
\Delta_\lambda\lambda^\alpha=\nu\lambda^\alpha\,,\quad
 \Delta_{\bar\lambda}\bar\lambda^\alpha=-\nu\bar\lambda^\alpha\,,
\end{align}
which satisfy
\begin{align}
(\nu-i\alpha(\sigma_0))(\bar\epsilon\gamma_\mu\lambda^{\alpha}+\epsilon\gamma_\mu\bar\lambda^{\alpha})
 -i\da_\mu(\bar\epsilon\lambda^{\alpha}-\epsilon\bar\lambda^{\alpha})
&=0\,,
\label{missing_vector_2d}
 \\
 (\nu-i\alpha(\sigma_0))(\bar\epsilon\gamma_3\lambda^{\alpha}+\epsilon\gamma_3\bar\lambda^{\alpha})
 -\alpha(\eta_0)(\bar\epsilon\lambda^{\alpha}-\epsilon\bar\lambda^{\alpha})
&=0\,.
\label{missing_sigma_2}
\end{align}
We can expand $\lambda$ and $\bar\lambda$ as
\begin{align}
\lambda^\alpha =\Lambda\bar\epsilon +\Lambda'\gamma^3\bar\epsilon
\,,\quad 
\bar\lambda^\alpha =\bar\Lambda\epsilon +\bar\Lambda'\gamma^3\epsilon
\,,
\end{align}
where $\Lambda,\Lambda',\bar\Lambda$ and $\bar\Lambda'$ are scalars.
Then the boundary conditions can be written as
\begin{align}
e^{-i\varphi}\Lambda|+e^{i\varphi}\bar\Lambda| =0
\,,\quad
e^{-i\varphi}\Lambda'|-e^{i\varphi}\bar\Lambda'| =0\,.
\label{2d_Lambda_bdry_cond}
\end{align}
The eigenvalue equation $\Delta_\lambda \lambda^\alpha= \nu \lambda^\alpha$ is equivalent to
\begin{align}
i v^\mu\da_\mu \Lambda +\tilde{v}^\mu\da_\mu \Lambda' 
+\frac{1}{\ell}(\Lambda+w\Lambda')
+\alpha(\eta_0)(w\Lambda-\Lambda')
&=-(\nu -i\alpha(\sigma_0))(\Lambda-w\Lambda')\,,
\label{eigen lambda1_2d}\\
\tilde{v}^\mu\da_\mu \Lambda +i v^\mu\da_\mu \Lambda' 
+\frac{1}{\ell}(w\Lambda+\Lambda')
+\alpha(\eta_0)(\Lambda-w\Lambda')
&=-(\nu -i\alpha(\sigma_0)) (w\Lambda-\Lambda')\,,
\label{eigen lambda2_2d}
\end{align}
and $\Delta_{\bar\lambda} \bar\lambda^\alpha= -\nu \bar\lambda^\alpha$ is equivalent to
\begin{align}
 \tilde{v}^\mu\da_\mu \bar\Lambda +i v^\mu\da_\mu \bar\Lambda' 
 +\frac{1}{\ell}(w\bar\Lambda-\bar\Lambda')
 -\alpha(\eta_0)(\bar\Lambda+w\bar\Lambda')
&=(\nu -i\alpha(\sigma_0)) (w\bar\Lambda+\bar\Lambda')\,,
\label{eigen barlambda1_2d}\\
 i v^\mu\da_\mu \bar\Lambda +\tilde{v}^\mu\da_\mu \bar\Lambda' 
 -\frac{1}{\ell}(\bar\Lambda-w\bar\Lambda')
 +\alpha(\eta_0)(w\bar\Lambda+\bar\Lambda')
&=-(\nu -i\alpha(\sigma_0)) (\bar\Lambda+w\bar\Lambda')\,.
\label{eigen barlambda2_2d}
\end{align}
We can rewrite equations \eqref{missing_vector_2d} and \eqref{missing_sigma_2}
into the following forms
\begin{align}
- (\nu -i\alpha(\sigma_0))[w(w_-\Lambda+w_+\bar\Lambda)-(w_-\Lambda'-w_+\bar\Lambda')]
-i v^\mu\da_\mu(w_-\Lambda'-w_+\bar\Lambda')
&=0\,,
\label{vA=0_2d}
\\
 -i(\nu -i\alpha(\sigma_0))[w_-\Lambda-w_+\bar\Lambda-w(w_-\Lambda'+w_+\bar\Lambda')]
-i \tilde{v}^\mu\da_\mu(w_-\Lambda'-w_+\bar\Lambda')
&=0\,,
\label{tildevA=0}
\\
 (\nu -i\alpha(\sigma_0))(w_-\Lambda+w_+\bar\Lambda)
-\alpha(\eta_0)(w_-\Lambda'-w_+\bar\Lambda')
&=0\,.
\label{sigma_2=0}
\end{align}
where $w_\pm$ is defined by \eqref{2d-bilinear}.
If we make the ansatz,
\begin{align}
\Lambda=f(\theta)e^{i (m+1)\varphi}\,,\quad
\Lambda'=g(\theta)e^{i (m+1)\varphi}\,,\quad
\bar\Lambda=\bar f(\theta)e^{i (m-1)\varphi}\,,\quad
\bar\Lambda'=\bar g(\theta)e^{i (m-1)\varphi}\,, 
\end{align}
where $m$ is an integer,
then \eqref{vA=0_2d}  and \eqref{sigma_2=0} can be written as
\begin{align}
 -\ell(\nu -i\alpha(\sigma_0))\cos\theta (f+\bar f)
 +[\ell(\nu -i\alpha(\sigma_0))+m-\ell\alpha(\eta_0)(\kappa-\cos\theta)](g-\bar g)
 &=0\,,
 \\
 \ell(\nu -i\alpha(\sigma_0)) (f+\bar f) -\ell\alpha(\eta_0)(g-\bar g)
 &=0\,.
\end{align}
These lead to
\begin{align}
[\ell(\nu-i\alpha(\sigma_0))+m-\kappa\,\ell\alpha(\eta_0)](g-\bar g)=0\,.
\end{align}
If $g-\bar g\neq 0$,\footnote{
We can see that,
if $g-\bar{g}=0$,
there are no consistent solutions 
with the regularity at $\theta=0$ and the boundary conditions at $\theta=\theta_0$,
except the case that $\nu =i\alpha(\sigma_0)$, $m=0$ and $\eta_0=0$.}
we have
\begin{align}
\ell\nu=i\ell \alpha(\sigma_0)-m+\kappa\,\ell\alpha(\eta_0)\,.
\end{align}
On the other hand, 
because \eqref{tildevA=0} should be consistent with 
\eqref{eigen lambda1_2d} and \eqref{eigen barlambda2_2d},
we obtain a relation
\begin{align}
(-m+\kappa\,\ell \alpha(\eta_0))(f-\bar f)
=\ell \alpha(\eta_0)(g+\bar g)\,.
\end{align}
Thus, we obtain the following relations:
\begin{align}
(-m+\kappa\,\ell \alpha(\eta_0))f
=\ell \alpha(\eta_0)g\,,
\label{f_and_g}
\\
(-m+\kappa\,\ell \alpha(\eta_0))\bar f
=-\ell \alpha(\eta_0)\bar g\,.
\label{bf_and_bg}
\end{align}
From the remaining conditions,
%we require that
$g(\theta)$ should satisfy
\begin{align}
\sin\theta\,\partial_\theta g+[(m+1-\kappa\,\ell\alpha(\eta_0))\cos\theta+ \ell\alpha(\eta_0)]g=0\,,
\end{align}
and
$\bar g(\theta)$ should satisfy
\begin{align}
\sin\theta\,\partial_\theta \bar g+[(-m+1+\kappa\,\ell\alpha(\eta_0))\cos\theta- \ell\alpha(\eta_0)]\bar g=0\,.
\end{align}
One can show that general solutions of these differential equations are given by
\begin{align}
g(\theta)&=C_1(-m+\kappa\,\ell\alpha(\eta_0))\Bigl(\sin\frac{\theta}{2}\Bigr)^{(\kappa-1)\ell\alpha(\eta_0)} 
\Bigl(\cos\frac{\theta}{2}\Bigr)^{(\kappa+1)\ell\alpha(\eta_0)} \,\sin^{-m-1}\theta\,,
\\
\bar g(\theta)&=C_2(-m+\kappa\,\ell\alpha(\eta_0))\Bigl(\sin\frac{\theta}{2}\Bigr)^{-(\kappa-1)\ell\alpha(\eta_0)} 
\Bigl(\cos\frac{\theta}{2}\Bigr)^{-(\kappa+1)\ell\alpha(\eta_0)} \,\sin^{m-1}\theta\,. 
\end{align}
In the case where $\theta_0<\pi$,
taking account of the regularity of fermionic eigenmodes at $\theta=0$
and the boundary condition \eqref{2d_Lambda_bdry_cond},
it is needed that $m=0$ and
\begin{align}
 C_1\,\Bigl(\cos\frac{\theta_0}{2}\Bigr)^{2\ell\alpha(\eta_0)} 
 -C_2\,\Bigl(\cos\frac{\theta_0}{2}\Bigr)^{-2\ell\alpha(\eta_0)}=0\,.
\end{align}
Therefore, the eigenvalue is given by
\begin{align}
\ell\nu=i\ell \alpha(\sigma_0)+\ell\alpha(\eta_0)\,.
\end{align}
For the special case that $\theta_0=\pi$,
there are the following solutions:
\begin{align}
 C_2&=0\,,\quad\nn\\
 \ell\nu&=i\ell \alpha(\sigma_0)-m+\ell\alpha(\eta_0)\,,\quad
\left\{ \begin{array}{ll}
m\leq 0 \,&\text{for}\quad 2\ell\alpha(\eta_0)\geq 2\\
m\leq 2\ell\alpha(\eta_0)-2\,&\text{for}\quad 2\ell\alpha(\eta_0)\leq 1\\
\end{array} \right.
\end{align}
or
\begin{align}
 C_1&=0\,,\quad\nn\\
  \ell\nu&=i\ell \alpha(\sigma_0)-m+\ell\alpha(\eta_0)\,,\quad
\left\{ \begin{array}{ll}
m\geq 0 \,&\text{for}\quad 2\ell\alpha(\eta_0)\leq-2\\
m>2\ell\alpha(\eta_0)+1\,&\text{for}\quad 2\ell\alpha(\eta_0)\geq -1\\
\end{array} \right.
\end{align}
In the case for $S^2$,
there are solutions:
\begin{align}
 C_2&=0\,,\quad\nn\\
 \ell\nu&=
\left\{ \begin{array}{ll}
i\ell \alpha(\sigma_0)-m+|\ell\alpha(\eta_0)|
&\text{for}\quad\ell\alpha(\eta_0)\neq 0\\
i\ell \alpha(\sigma_0)-m+1
&\text{for}\quad\ell\alpha(\eta_0)= 0\\
\end{array} \right.
\quad (m\leq 0)
\end{align}
or
\begin{align}
 C_1&=0\,,\quad\nn\\
 \ell\nu&=
\left\{ \begin{array}{ll}
i\ell \alpha(\sigma_0)-m-|\ell\alpha(\eta_0)|
&\text{for}\quad\ell\alpha(\eta_0)\neq 0\\
i\ell \alpha(\sigma_0)-m-1
&\text{for}\quad\ell\alpha(\eta_0)= 0\\
\end{array} \right.
\quad (m\geq 0)
\end{align}

Therefore, 
up to an overall constant,
the 1-loop factor for the vector multiplet is given by
\begin{align}
 Z^{\rm{1-loop}}_{\rm{vector}} &=
 \prod_{\alpha\in\Delta_+}
 (i \alpha(\sigma_0)+\alpha(\eta_0))\,,
\end{align}
for $\theta_0<\pi$\,.
For $\theta_0=\pi$,
the 1-loop factor is
\begin{eqnarray}
 Z^{\rm{1-loop}}_{\rm{vector}} &=&
\prod_{\alpha\in\Delta_+}
\prod_{ \{ \ell\alpha(\eta_0),m  \} \in D_f} \,
(i\ell\alpha(\sigma_0)-m+\ell\alpha(\eta_0))\nn\\
&&\times
\prod_{\alpha\in\Delta_+}
\prod_{ \{ \ell\alpha(\eta_0),m  \} \in D_b} \,
\,(i\ell\alpha(\sigma_0)-m+\ell\alpha(\eta_0))^{-1}
\nn\\
&=& \prod_{\alpha\in\Delta_+,\, |2\ell\alpha(\eta_0)| \geq 2}\,
(i\ell\alpha(\sigma_0)+\ell\alpha(\eta_0))
%\prod_{\alpha\in\Delta_+,\,2\ell\alpha(\eta_0)\leq-2}\,
%(i\ell\alpha(\sigma_0)+\ell\alpha(\eta_0))
\,,
\end{eqnarray}
where
\begin{eqnarray}
 D_f=\{ 
\{ 2\ell\alpha(\eta_0)\geq 2 , m\leq 0 \, \} \cup  
\{ 2\ell\alpha(\eta_0)\leq 1, m\leq 2\ell\alpha(\eta_0)-2 \, \} \nn \\
\cup  
\{ 2\ell\alpha(\eta_0)\leq -2, m\geq 0 \, \} \cup  
\{ 2\ell\alpha(\eta_0)\geq -1, m\geq 2\ell\alpha(\eta_0)+2 \, \}
\},
\end{eqnarray}
and
\begin{eqnarray}
 D_b=\{ 
\{ 2\ell\alpha(\eta_0)\geq 1 , m\leq -1 \, \} \cup  
\{ 2\ell\alpha(\eta_0)\leq 0, m\leq 2\ell\alpha(\eta_0)-2 \, \} \nn \\
\cup  
\{ 2\ell\alpha(\eta_0)\leq -1, m\geq 1 \, \} \cup  
\{ 2\ell\alpha(\eta_0)\geq 0, m\geq 2\ell\alpha(\eta_0)+2 \, \}
\}.
\end{eqnarray}
%\begin{align}
% Z^{\rm{1-loop}}_{\rm{vector}} &=
%\prod_{\alpha\in\Delta_+,\, 2\ell\alpha(\eta_0)\geq 2}\prod_{m\leq 0}\,(i\ell\alpha(\sigma_0)-m+\ell\alpha(\eta_0))
%\prod_{\alpha\in\Delta_+,\, 2\ell\alpha(\eta_0)\leq 1}\prod_{m\leq 2\ell\alpha(\eta_0)-2}\,(i\ell\alpha(\sigma_0)-m+\ell\alpha(\eta_0))
%\nn\\
%&
%\prod_{\alpha\in\Delta_+,\, 2\ell\alpha(\eta_0)\leq-2}\prod_{m\geq 0}\,(i\ell\alpha(\sigma_0)-m+\ell\alpha(\eta_0))
%\prod_{\alpha\in\Delta_+,\, 2\ell\alpha(\eta_0)\geq-1}\prod_{m\geq 2\ell\alpha(\eta_0)+2}\,(i\ell\alpha(\sigma_0)-m+\ell\alpha(\eta_0))
%\nn\\
%&
%\prod_{\alpha\in\Delta_+,\, 2\ell\alpha(\eta_0)\geq0}\prod_{m\geq 2\ell\alpha(\eta_0)+2}\,(i\ell\alpha(\sigma_0)-m+\ell\alpha(\eta_0))^{-1}
%\prod_{\alpha\in\Delta_+,\, 2\ell\alpha(\eta_0)\leq-1}\prod_{m\geq 1}\,(i\ell\alpha(\sigma_0)-m+\ell\alpha(\eta_0))^{-1}
%\nn\\
%&
%\prod_{\alpha\in\Delta_+,\, 2\ell\alpha(\eta_0)\geq 1}\prod_{m\leq -1}\,(i\ell\alpha(\sigma_0)-m+\ell\alpha(\eta_0))^{-1}
%\prod_{\alpha\in\Delta_+,\, 2\ell\alpha(\eta_0)\leq 0}\prod_{m\leq  2\ell\alpha(\eta_0)-2}\,(i\ell\alpha(\sigma_0)-m+\ell\alpha(\eta_0))^{-1}
%\nn\\
%&=\prod_{\alpha\in\Delta_+,\, |\ell\alpha(\eta_0)| \geq 1}\,
%(i\ell\alpha(\sigma_0)+\ell\alpha(\eta_0))
%%\prod_{\alpha\in\Delta_+,\,2\ell\alpha(\eta_0)\leq-2}\,
%%(i\ell\alpha(\sigma_0)+\ell\alpha(\eta_0))
%\,,
%\end{align}

For $S^2$, the 1-loop factor is given by
\begin{eqnarray}
Z^{\rm{1-loop}}_{\rm{vector}} &=&
 \prod_{\alpha\in\Delta_+,\, \alpha(\eta_0)=0}\prod_{m\geq 0}\,
 (\ell^2\alpha(\sigma_0)^2 +(m+1)^2) 
%\nn \\
%&& 
\prod_{\alpha\in\Delta_+,\, \alpha(\eta_0)\neq0 }\prod_{m\geq 0}\,
 (\ell^2\alpha(\sigma_0)^2 +(m+|\ell\alpha(\eta_0)|)^2) \nn \\
&&
\times \prod_{\alpha\in\Delta_+}\prod_{m\geq1}(\ell^2\alpha(\sigma_0)^2 +(m+|\ell\alpha(\eta_0)|)^2)^{-1}
\nn\\
&=& \prod_{\alpha\in\Delta_+,\,\, \alpha(\eta_0)\neq0}
 (\ell^2\alpha(\sigma_0)^2 +\ell^2\alpha(\eta_0)^2)\,,
\end{eqnarray}
The result for $S^2$ is same as \cite{Benini:2012ui, Doroud:2012xw}.

%%%%%%%%%%%%%%%%%%%%%%%%%%%%%%%%%%%%%%%%%%%%%%%%%%%%%%%%%%%%%
\subsubsection{Chiral multiplet}
Next let us consider the 1-loop determinant for the chiral multiplet.

Expanding fields around the saddle point and leaving only the quadratic terms,
we have
\begin{align}
t \int \rmd^2 x \sqrt{g}\, \delta V_{\mathrm{chiral}} &=
 \int \rmd^2 x \sqrt{g}\, \mathcal{L}_{\text{reg}}
 +\mathcal{O}(t^{-1/2})\,,
\end{align}
where
\begin{align}
 \mathcal{L}_{\text{reg}} &=
\bar\phi\,
\Delta_\phi\,
\phi 
+\bar\psi\,
\Delta_\psi\,
\psi
\,,
\\
\Delta_\phi &=
-\da_\mu D^{(a)\mu} +\sigma_0^2 +\eta_0^2+i\frac{q-1}{\ell}\sigma_0-\frac{q(q-2)}{4\ell^2}\,,
\\
\Delta_\psi &=
-i\gamma^\mu\da_\mu +i\sigma_0 -\eta_0\gamma^3-\frac{q}{2\ell}\,.
\end{align}
Hence, to evaluate the 1-loop determinant,
we consider the eigenvalue problems for $\Delta_\phi$ and $\Delta_\psi$.
Hereafter, we set $\beta = i\sigma_0 -(q-1)/2\ell$\,.

We can see that there are partial cancellations in the 1-loop factor between the
contributions
from the bosonic 
and fermionic eigenmodes.
Let $\psi$ be a fermionic eigenmode:
$\Delta_\psi\psi=\nu\,\psi$.
Then, if we define $\phi_1\equiv \bar\epsilon\psi$,
we find that $\phi_1$ is a scalar eigenmode :
$\Delta_\phi \phi_1=\nu(\nu-2\beta)\,\phi_1$\,.
On the other hand, using a scalar eigenmode
($\Delta_\phi \phi =M^2\phi$),
we define
\begin{align}
\psi_\pm \equiv \Bigl(\nu_\pm-\beta+\frac{1}{2\ell}\Bigr)\,\epsilon\phi
-i\gamma^\mu \epsilon\, \da_\mu\phi -\eta_0\gamma^3\epsilon\phi\,,
\end{align}
where $\nu_\pm\equiv\beta\pm\sqrt{M^2+\beta^2}$\,.
Then, we find that
\begin{align}
\Delta_\psi \psi_\pm =
\nu_\pm\,\psi_\pm
\,.
\end{align}
Note that $\phi_1$ and $\psi_\pm$ satisfy the boundary conditions \eqref{2d_chiral_bc}.

%%%%%%%%%%%%%%%%%%%%%%%%%%%%%%%%%%%%%%%%%%%%%%%%%%%%%%%%%%%%%
\paragraph{Unpaired eigenmodes}~\\

We consider the unpaired fermionic eigenmode.
If $\phi_1(\equiv\bar\epsilon\psi)=0$, 
$\psi$ can be written as $\psi=\bar\epsilon\Psi$,
where $\Psi$ is a scalar function on which 
any boundary condition is not imposed.
Since $\psi$ is a fermionic eigenmode, $\Delta_\psi\psi=\nu\,\psi$,
we have 
\begin{align}
i v^\mu\da_\mu \Psi &=\Bigl(\nu-\beta-\frac{1}{2\ell}-w\,\eta_0\Bigr)\Psi\,,\\
\tilde{v}^\mu\da_\mu\Psi &=\Bigl[w\Bigl(\nu-\beta-\frac{1}{2\ell}\Bigr)-\eta_0\Bigr]\Psi\,.
\end{align}
Therefore, we obtain the solutions:
\begin{align}
 \Psi &\propto\Bigl(\sin\frac{\theta}{2}\Bigr)^{(\kappa-1)\ell\eta_0} 
\Bigl(\cos\frac{\theta}{2}\Bigr)^{(\kappa+1)\ell\eta_0} \,\sin^{m}\theta\,
e^{-i m\varphi}\,,
\end{align}
where $m$ is an integer.
The corresponding eigenvalues for weights $\rho$ are given by
\begin{align}
 \nu &=i\rho(\sigma_0)-\frac{q-2}{2\ell}+\rho(\eta_0)+\frac{m}{\ell}
\,\quad (m\geq 0)\,,
\end{align}
in the case where $\theta_0<\pi$\,,
and
\begin{align}
 \nu &=i\rho(\sigma_0)-\frac{q-2}{2\ell}+\rho(\eta_0)+\frac{m}{\ell}\,,
\quad\left\{ \begin{array}{ll}
m\geq 0 \,&\text{for}\quad\ell\rho(\eta_0)\geq 0\\
m\geq -2\,\ell\rho(\eta_0)\,&\text{for}\quad\ell\rho(\eta_0)\leq 0\\
\end{array} \right.\,,
\end{align}
in the case where $\theta_0=\pi$\,,
and
\begin{align}
 \nu &=i\rho(\sigma_0)-\frac{q-2}{2\ell}+|\rho(\eta_0)|+\frac{m}{\ell}
\,\quad (m\geq 0)\,,
\end{align}
for $S^2$\,.

On the other hand,
the unpaired bosonic eigenmodes, $\Delta_\phi \phi =M^2\phi$,
should satisfy 
\begin{align}
 \Bigl(\nu-\beta+\frac{1}{2\ell}\Bigr)\,\epsilon\phi
-i\gamma^\mu \epsilon\, \da_\mu\phi -\eta_0\gamma^3\epsilon\phi=0 
\quad\quad (\nu(\nu-2\beta)=M^2) \,,
\end{align} 
which is equivalent to
\begin{align}
i v^\mu\da_\mu \phi &=\Bigl(\nu-\beta+\frac{1}{2\ell}-w\,\eta_0\Bigr)\phi\,,
\\
\tilde{v}^\mu\da_\mu\phi &=-\Bigl[w\Bigl(\nu-\beta+\frac{1}{2\ell}\Bigr)-\eta_0\Bigr]\phi\,.
\label{2d_chiral_unpaired_scalar} 
\end{align}
It can be easily checked that
these equations lead to $\Delta_\phi \phi =M^2\phi$\,.
In the case where $\theta_0<\pi$\,,
there is no nontrivial solution which satisfies 
equation \eqref{2d_chiral_unpaired_scalar} 
and the boundary condition $\phi|=0$ simultaneously.
In the case where $\theta_0=\pi$ and the case for $S^2$\,,
we obtain the solutions
\begin{align}
 \phi &\propto\Bigl(\sin\frac{\theta}{2}\Bigr)^{-(\kappa-1)\ell\eta_0} 
\Bigl(\cos\frac{\theta}{2}\Bigr)^{-(\kappa+1)\ell\eta_0} \,\sin^{m}\theta\,
e^{i m\varphi}\,.
\end{align}
The corresponding eigenvalues are given by
\begin{align}
 \nu &=i\rho(\sigma_0)-\frac{q}{2\ell}+\rho(\eta_0)-\frac{m}{\ell}\,,
\quad\left\{ \begin{array}{ll}
m\geq 2\,\ell\rho(\eta_0)+1 \,&\text{for}\quad\ell\rho(\eta_0)\geq 0\\
m\geq 0\,&\text{for}\quad\ell\rho(\eta_0)< 0\\
\end{array} \right.\,,
\end{align}
for $\theta_0=\pi$\,,
and
\begin{align}
 \nu &=i\rho(\sigma_0)-\frac{q}{2\ell}-|\rho(\eta_0)|-\frac{m}{\ell}
\,\quad (m\geq 0)\,,
\end{align}
for $S^2$\,.

Therefore, 
up to an overall constant,
the 1-loop determinant for the chiral multiplet is given by
\begin{align}
 Z^{\rm{1-loop}}_{\rm{chiral}} &=
 \prod_{\rho}\prod_{m\geq 0}\, \Bigl(i\ell\rho(\sigma_0)-\frac{q}{2}+\ell\rho(\eta_0)+m+1 \Bigr)\,,
\end{align}
for $\theta_0<\pi$.
For the special case that $\theta_0=\pi$,
we obtain\footnote{
$\rho_+$ (or $\rho_-$) means the set of weight vectors such that
$\rho(\eta_0)\geq 0$ (or $\rho(\eta_0)< 0$).} 
\begin{align}
 Z^{\rm{1-loop}}_{\rm{chiral}} &=\prod_{\rho\in \rho_+}\Biggl(
 \frac{\prod_{m\geq 0}\, \Bigl(i\ell\rho(\sigma_0)-\frac{q}{2}+\ell\rho(\eta_0)+m+1\Bigr)}
{\prod_{m\geq 2\ell\rho(\eta_0) +1}\,\Bigl(-i\ell\rho(\sigma_0)+\frac{q}{2}-\ell\rho(\eta_0)+m\Bigr)}\Biggr)
\nn\\
&\quad\,\times\prod_{\rho\in \rho_-}\Biggl(
 \frac{\prod_{m\geq  -2\ell\rho(\eta_0)}\, \Bigl(i\ell\rho(\sigma_0)-\frac{q}{2}+\ell\rho(\eta_0)+m+1\Bigr)}
{\prod_{m\geq 0}\,\Bigl(-i\ell\rho(\sigma_0)+\frac{q}{2}-\ell\rho(\eta_0)+m\Bigr)}\Biggr)\,.
\end{align}
For $S^2$, 
the 1-loop factor is given by
\begin{align}
Z^{\rm{1-loop}}_{\rm{chiral}} &=\prod_{\rho}
  \frac{\prod_{m\geq 0}\, \Bigl(i\ell\rho(\sigma_0)-\frac{q}{2}+|\ell\rho(\eta_0)|+m+1\Bigr)}
{\prod_{m\geq 0}\,\Bigl(-i\ell\rho(\sigma_0)+\frac{q}{2}+|\ell\rho(\eta_0)|+m\Bigr)}\,.
\end{align}
The result for $S^2$ is same as \cite{Benini:2012ui, Doroud:2012xw}.

%%%%%%%%%%%%%%%%%%%%%%%%%%%%%%%%%%%%%%%%%%%%%%%%%%%%%%%%%%%%%
\subsection{Partition functions and Wilson loops}

From what we have obtained,
%We summarize the obtained results.
we find that the exact partition function 
for $\theta_0 <\pi$
is given by
\begin{eqnarray}
 Z= Z_{\mathrm{classical}} \, 
Z^{\rm{1-loop}}_{\rm{vector}} \, Z^{\rm{1-loop}}_{\rm{chiral}}, 
\end{eqnarray}
where 
\begin{eqnarray}
 Z_{\mathrm{classical}} =e^{-i(\frac{\zeta}{\ell} \tr \,\sigma_0
+\frac{\Theta}{ 2\pi \ell} \tr\, \eta_0)V(\theta_0)},
\end{eqnarray}
where $\sigma_0$ and $\eta_0$ were fixed at the boundary
and $V(\theta_0)$ is the volume of the manifold we consider: $V(\theta_0)=2\pi(1-\cos\theta_0)\ell^2$.

The supersymmetric Wilson loop operator is given by the following form
\begin{align}
 W_R = \frac{1}{\rm{dim}\,R}
 \tr_R \,\mathrm{P} \exp
 \biggl( \oint_{\theta=\theta_1} \rmd\varphi 
(i A_\varphi +\ell(-\sigma_1+i\cos\theta\sigma_2)\biggr)\,,
\end{align}
where $R$ is a representation of the gauge group,
and $\mathrm{P}$ represents path-ordering
and its path is given by $\theta=\theta_1, (0<\theta_1<\theta_0)$.
This operator is actually invariant under the supersymmetry transformation
generated by the Killing spinors \eqref{bks2}.
Thus, 
we find that the expectation value of the supersymmetric Wilson loop is exactly
\begin{align}
 \langle W_R \rangle = \frac{1}{\rm{dim}\,R}
\tr_R \exp\biggl( 2 \pi 
(i \ell\eta_0 -\ell\sigma_0)\biggr)\,.
\end{align}
%%%%%%%%%%%%%%%%%%%%%%%%%%%%%%%%%%%%%%%%%%%%%%%%%%%%%%%%%%%%%

\bigskip
%\goodbreak
\centerline{\bf Acknowledgments}
%\acknowledgments
%\noindent
We would like to thank especially 
K. Hosomichi for helpful discussions on many points in the paper
and
K. Sakai for %helpful discussions and 
collaboration at the early stage of this work. 
We would also 
like to thank N. Hama and T. Nosaka for helpful discussions.
S.T. was supported in part by JSPS KAKENHI Grant Number 23740189. 

%%%%%%%%%%%%%%%%%%%%%%%%%%%%%%%%%%%%%%%%%%%%%%%%%%%%%%%%%%%%%
%%%%%%%%%%%%%%%%%%%%%%%%%%%%%%%%%%%%%%%%%%%%%%%%%%%%%%%%%%%%%
\appendix
%%%%%%%%%%%%%%%%%%%%%%%%%%%%%%%%%%%%%%%%%%%%%%%%%%%%%%%%%%%%%
\section{Notations and useful formulas}
\label{notation}

In this Appendix, we will explain the notations
used in the paper and summarize some useful formulas.

\paragraph{Indices}

We use the following conventions for indices:
\begin{eqnarray*}
\hbox{coordinate indices} \qquad \mu,\nu,\cdots 
\\
\hbox{tangent space indices} \qquad  a,b,\cdots
\\
\hbox{spinor indices} \qquad \alpha,\beta,\cdots
\end{eqnarray*}

\paragraph{Gamma matrices}
In this paper, we take gamma matrices  as
\begin{align}
\gamma^1=
\begin{pmatrix}
0&1\\
1&0
\end{pmatrix}
\,,\,
\gamma^2=
\begin{pmatrix}
0&-i\\
i&0
\end{pmatrix}
\,,\,
\gamma^3=
\begin{pmatrix}
1&0\\
0&-1
\end{pmatrix}
\,. 
\end{align}

\paragraph{Spinors}
For both the three and two dimensional theories, 
we use the two-components Dirac spinors.
The spinor-bi-linears are defined as  
\begin{align}
\bar{\epsilon}\lambda 
&\equiv \bar{\epsilon}_{\alpha}\lambda^{\alpha}
\equiv C_{\alpha\beta}\bar{\epsilon}^{\alpha}\lambda^{\beta}\,,\\
\bar{\epsilon}\gamma^a\lambda 
&\equiv \bar{\epsilon}_{\alpha}(\gamma^{a})^{\alpha}_{~\beta}\lambda^{\beta}\,,
\end{align}
where
$C_{\alpha\beta}$ is the antisymmetric matrix ($C_{12}=-C_{21}=1$).
It is easy to check that
\begin{align}
 \bar{\epsilon}\lambda = \lambda\bar{\epsilon}\,,\quad
 \bar{\epsilon}\gamma^a\lambda =-\lambda\gamma^a\bar{\epsilon}\,,
\end{align}
for Grassmann odd spinors.

\paragraph{Useful formulas in 3D}
\label{useful_3d}
The Fierz identity for Grassmann odd spinors in 3D:
\begin{align}
 (\eta\lambda)(\epsilon\psi)
 =-\frac{1}{2}(\eta\epsilon)(\lambda\psi)
  +\frac{1}{2}(\eta\gamma^\mu\epsilon)(\lambda\gamma_\mu\psi)\,.
\end{align}
Grassmann even positive Killing spinors on $S^3$ are spanned by
\begin{align}
\epsilon = \frac{1}{\sqrt{2}}
\begin{pmatrix}
-e^{-\frac{i}{2}(\varphi-\chi-\theta)}\\
e^{-\frac{i}{2}(\varphi-\chi+\theta)}
\end{pmatrix} 
\quad 
\text{and} 
\quad
\bar\epsilon =
\frac{1}{\sqrt{2}}
\begin{pmatrix}
e^{\frac{i}{2}(\varphi-\chi+\theta)}\\
e^{\frac{i}{2}(\varphi-\chi-\theta)}
\end{pmatrix}
\,.
\end{align}
The bi-linears of these Grassmann even spinors are given by
\begin{align}
%\begin{split}
\bar\epsilon \epsilon &=1 \,,\quad
v^a \equiv\bar\epsilon\gamma^a \epsilon = (-\cos\theta, \sin\theta, 0)\,,
\\
v^a_+&\equiv\epsilon\gamma^a \epsilon = (i\sin\theta, i\cos\theta, 1)e^{-i(\varphi-\chi)}\,,
\\
v^a_-&\equiv\bar\epsilon\gamma^a \bar\epsilon = (i\sin\theta, i\cos\theta, -1)e^{+i(\varphi-\chi)}\,,
%\end{split}
\end{align}
We summarize some useful formulas for the above Killing spinors and Killing vectors:
\begin{align}
%\begin{split}
v_\mu\gamma^\mu\epsilon &=\epsilon \,,
\quad 
v^+_\mu\gamma^\mu\epsilon =0 \,,
\quad 
v^-_\mu\gamma^\mu\epsilon =2\bar\epsilon \,,
\\
v_\mu\bar\epsilon\gamma^\mu&=\bar\epsilon \,,
\quad 
v^+_\mu\bar\epsilon\gamma^\mu=2\epsilon \,,
\quad 
v^-_\mu\bar\epsilon\gamma^\mu=0 \,,
\\
v^\mu v_\mu&=1 \,,
\quad
v^\mu_+ v^-_\mu = -2 \,,
\quad
\varepsilon_{abc}v^b v^c_{\pm} = \mp i \,v^\pm_a \,,
\quad
\varepsilon_{abc}v_+^b v_-^c = 2 i \,v_a \,,
\\
D_\mu v^X_\nu &= \frac{1}{\ell}\varepsilon_{\mu\nu\rho}v^\rho_X 
\qquad (X= \text{no mark}, +, -)\,,
\\
g^{\mu \nu}&= v^\mu v^\nu -\frac{1}{2}(v_+^\mu v_-^\nu + v_-^\mu v_+^\nu)\,,
\end{align}
and formulas for an arbitrary scalar function $Y$:
\begin{align}
[v_+^\mu \da_\mu, v_-^\nu \da_\nu] Y&=-\frac{4i}{\ell}v^\mu \da_\mu Y\,,
\qquad
[v^\mu \da_\mu, v_{\pm}^\nu \da_\nu] Y=\pm\frac{2i}{\ell}v_\pm^\mu \da_\mu Y\,,
\\
D^{(a)\mu} \da_\mu Y
&=v^\mu \da_\mu (v^\nu \da_\nu Y)
-\frac{1}{2} v_+^\mu \da_\mu (v_-^\nu \da_\nu Y)
-\frac{1}{2} v_-^\mu \da_\mu (v_+^\nu \da_\nu Y)
\\
&=v^\mu \da_\mu (v^\nu \da_\nu Y)
-v_-^\mu \da_\mu (v_+^\nu \da_\nu Y)
+\frac{2i}{\ell} v^\mu \da_\mu Y\,,
%\end{split}
\end{align}
where 
$\da_\mu Y\equiv(\nabla_\mu -i a_\mu )Y$ 
and $a_\mu$ is a background gauge field which satisfies 
$F^{(a)}_{\mu\nu}\equiv \nabla_\mu a_\nu -\nabla_\nu a_\mu -i[a_\mu,a_\nu]=0$ .

\paragraph{Useful formulas in 2D}
\label{useful_2d}

The Fierz identity for Grassmann odd spinors in 2D is given by
\begin{align}
 (\eta\lambda)(\epsilon\psi)
 =-\frac{1}{2}(\eta\epsilon)(\lambda\psi)
  +\frac{1}{2}(\eta\gamma^3\epsilon)(\lambda\gamma_3\psi)
  +\frac{1}{2}(\eta\gamma^\mu\epsilon)(\lambda\gamma_\mu\psi)\,.
\end{align}

Any Grassmann even positive Killing spinor on $S^2$ can be spanned by
\begin{align}
\epsilon =e^{i\frac{\varphi}{2}}
\begin{pmatrix}
i\cos\frac{\theta}{2}\\
-\sin\frac{\theta}{2}
\end{pmatrix}  \qquad \text{and} \qquad
\bar{\epsilon} =e^{-i\frac{\varphi}{2}}
\begin{pmatrix}
-\sin\frac{\theta}{2}\\
i\cos\frac{\theta}{2}
\end{pmatrix}\,.
\end{align}
We can compute the bi-linears of these Grassmann even spinors:
\begin{align}
\begin{split}
\label{2d-bilinear}
\bar\epsilon\epsilon &= 1\,,\quad
w \equiv\bar\epsilon\gamma^3\epsilon =\cos\theta  \,,\\
v^\mu&\equiv\bar\epsilon\gamma^\mu\epsilon =(0,1/\ell) \,,\quad
\tilde{v}^\mu\equiv\varepsilon^{\mu\nu}v_\nu=(\sin\theta/\ell,0)\,,
\\
v^\mu_+&\equiv\epsilon \gamma^\mu \epsilon=-\frac{e^{i\varphi}}{\sin\theta}(\tilde{v}^\mu+i \,w\, v^\mu)\,,\quad
v^\mu_-\equiv\bar\epsilon \gamma^\mu \bar\epsilon=\frac{e^{-i\varphi}}{\sin\theta}(\tilde{v}^\mu-i \,w\, v^\mu)\,,
\\
w_+&\equiv\epsilon \gamma^3 \epsilon=i\,e^{i\varphi}\sin\theta\,,\quad
w_-\equiv\bar\epsilon \gamma^3 \bar\epsilon=i\,e^{-i\varphi}\sin\theta\,.
\end{split}
\end{align}
We summarize some useful formulas for the above Killing spinors and their bi-linears:
\begin{align}
\begin{split}
v^\mu v_\mu &=\tilde{v}^\mu \tilde{v}_\mu=\sin^2\theta=1-w^2\,,\quad
v^\mu_{\pm}v_\mu=-w\,w_{\pm}\,,\quad
v^\mu_{\pm}\tilde{v}_\mu=\pm i\,w_{\pm}\,,
\\
 D_\mu v_\nu &= \frac{1}{\ell}\varepsilon_{\mu\nu}w\,,\quad
 D_\mu\tilde{v}_\nu=\frac{1}{\ell}g_{\mu\nu}w\,,\quad
 D_\mu w=-\frac{1}{\ell}\varepsilon_{\mu\nu}v^\nu=-\frac{1}{\ell}\tilde{v}_\mu\,,
\\
 v^\mu\gamma_\mu\epsilon+w\gamma^3\epsilon&=\epsilon\,,\,\,\,
v^\mu\gamma_\mu\bar\epsilon+w\gamma^3\bar\epsilon=-\bar\epsilon\,,\,\,\,
\tilde{v}^\mu\gamma_\mu\epsilon=-i w_+\bar\epsilon\,,\,\,\,
\tilde{v}^\mu\gamma_\mu\bar\epsilon=-i w_-\epsilon\,,\,\,\,
\\
\ast\da\ast\da Y&=\frac{1}{1-w^2} [v^\mu\da_\mu (v^\nu\da_\nu Y)
+\tilde{v}^\mu\da_\mu (\tilde{v}^\nu\da_\nu Y)]\,,
\end{split}
\end{align}
where $Y$ is an arbitrary scalar function and 
$\da_\mu Y\equiv(\nabla_\mu -i a_\mu )Y$ 
and $a_\mu$ is a background gauge field.
%%%%%%%%%%%%%%%%%%%%%%%%%%%%%%%%%%%%%%%%%%%%%%%%%%%%%%%%%%%%%
\section{Supersymmetry variations}
\label{susy-vari}
In this Appendix,
we confirm that the supersymmetry variations of actions
can be written by surface terms which vanish 
by imposing the boundary conditions.
	
%%%%%%%%%%%%%%%%%%%%%%%%%%%%%%%%%%%%%%%%%%%%%%%%%%%%%%%%%%%%%
\paragraph{Three-dimensional theories}
We summarize the variation of the actions 
under the supersymmetry transformations \eqref{susytr1}-\eqref{susytr2},
where we assume that SUSY parameters, 
$\epsilon, \bar\epsilon$, are positive Killing spinors.

Yang-Mills Lagrangian:
\begin{align}
\delta \mathcal{L}_{\mathrm{YM}}&=
\frac{1}{4}\tr\, D_\mu\Bigl[
\frac{1}{2}\varepsilon^{\mu\nu\rho}F_{\nu\rho}(\bar\lambda\epsilon+\lambda\bar\epsilon)
+D_\nu\sigma(\bar\lambda\gamma^{\mu\nu}\epsilon-\lambda\gamma^{\mu\nu}\bar\epsilon)
\nn\\
&\qquad\qquad\quad
+i F^{\mu\nu}(\bar\lambda\gamma_{\nu}\epsilon+\lambda\gamma_{\nu}\bar\epsilon)
- D^\mu\sigma(\bar\lambda\epsilon-\lambda\bar\epsilon)
- i(D+\sigma/\ell)(\bar\lambda\gamma^{\mu}\epsilon-\lambda\gamma^{\mu}\bar\epsilon)\Bigr]\,.
\end{align}

Chern-Simons term:
\begin{align}
\delta \mathcal{L}_{\mathrm{CS}}&=-\frac{i}{2}\tr\, D_\mu[
 \varepsilon^{\mu\nu\rho}A_\nu(\bar\lambda\gamma_{\rho}\epsilon+\lambda\gamma_{\rho}\bar\epsilon)
 +2\sigma(\bar\lambda\gamma^{\mu}\epsilon-\lambda\gamma^{\mu}\bar\epsilon)]\,.
\end{align}
	
FI term:
\begin{align}
\delta \mathcal{L}_{\mathrm{FI}}&=
\frac{\zeta}{2\pi\ell} \tr\, D_\mu(\bar\lambda\gamma^{\mu}\epsilon-\lambda\gamma^{\mu}\bar\epsilon)\,.
\end{align}

The matter kinetic terms:
\begin{align}
\delta \mathcal{L}_{\mathrm{mat}}&= \frac{1}{2} D_\mu\Bigl(
\bar\psi\epsilon D^\mu\phi-\bar\psi\gamma^{\mu\nu}\epsilon D_\nu\phi
+i\bar{F}\epsilon\gamma^\mu\psi
+\bar\phi\bar\lambda\gamma^{\mu}\epsilon\phi
-\bar\psi\gamma^\mu\epsilon\,\sigma\phi-i\frac{q}{\ell}\bar\psi\gamma^\mu\epsilon\,\phi
\nn\\
&\qquad\quad\,\,\,\,\,
+D^\mu\bar\phi\,\bar\epsilon\psi+D_\nu\bar\phi\,\bar\epsilon\gamma^{\mu\nu}\psi
-i\bar\psi\gamma^\mu\bar\epsilon\, F
-\bar\phi\lambda\gamma^{\mu}\bar\epsilon\phi
+\bar\phi\,\sigma\bar\epsilon\gamma^\mu\psi
+i\frac{q}{\ell}\bar\phi\,\bar\epsilon\gamma^\mu\psi
\Bigr)\,.
\end{align}

Therefore, the supersymmetry variations of actions
can be written by surface terms, 
and actually they vanish if we assume that the Killing spinors satisfy
the relations \eqref{3d_e_be} and \eqref{3d_e_be 2} 
and the fields satisfy the boundary conditions \eqref{3d-bc-a}-\eqref{3d-bc-la} and \eqref{3d_chiral_bc}. 

%%%%%%%%%%%%%%%%%%%%%%%%%%%%%%%%%%%%%%%%%%%%%%%%%%%%%%%%%%%%%
\paragraph{Two-dimensional theories}
We summarize the variation of the actions 
under the supersymmetry transformations \eqref{2d-vector_susy_trsf}-\eqref{2d-chiral_susy_trsf},
where we assume that SUSY parameters, 
$\epsilon$ and $\bar\epsilon$, are positive Killing spinors.

Yang-Mills Lagrangian:
\begin{align}
\delta \mathcal{L}_{\mathrm{YM}}&=
\frac{1}{4}\tr\, D_\mu[
-(F_{12}-\sigma_2/\ell)(\bar\lambda\gamma^\mu\gamma^3\epsilon+\lambda\gamma^\mu\gamma^3\bar\epsilon)
+i[\sigma_1,\sigma_2] (\bar\lambda\gamma^{\mu}\gamma^3\epsilon-\lambda\gamma^{\mu}\gamma^3\bar\epsilon)
\nn\\
&\qquad\qquad\quad
+D_\nu\sigma_1(\bar\lambda\gamma^{\mu\nu}\epsilon-\lambda\gamma^{\mu\nu}\bar\epsilon)
-i D_\nu\sigma_2(\bar\lambda\gamma^{\mu\nu}\gamma^3\epsilon+\lambda\gamma^{\mu\nu}\gamma^3\bar\epsilon)
\nn\\
&\qquad\qquad\quad
- D^\mu\sigma_1(\bar\lambda\epsilon-\lambda\bar\epsilon)
+i D^\mu\sigma_2(\bar\lambda\gamma^3\epsilon-\lambda\gamma^3\bar\epsilon)
-i(D+\sigma_1/\ell)(\bar\lambda\gamma^{\mu}\epsilon-\lambda\gamma^{\mu}\bar\epsilon)]\,.
\end{align}

FI term:
\begin{align}
\delta \mathcal{L}_{\mathrm{FI}}&=-\tr\, D_\mu\Bigl[
\frac{\zeta}{2}(\bar\lambda\gamma^{\mu}\epsilon-\lambda\gamma^{\mu}\bar\epsilon)
+\frac{\Theta}{4\pi}\varepsilon^{\mu\nu}(\bar\lambda\gamma_{\nu}\epsilon+\lambda\gamma_{\nu}\bar\epsilon)
\Bigr]\,.
\end{align}

The matter kinetic terms:
\begin{align}
\delta \mathcal{L}_{\mathrm{mat}}&= \frac{1}{2} D_\mu\Bigl(
\bar\psi\epsilon D^\mu\phi-\bar\psi\gamma^{\mu\nu}\epsilon D_\nu\phi
+i\bar{F}\epsilon\gamma^\mu\psi
\nn\\
&\qquad\qquad
+\bar\phi\bar\lambda\gamma^{\mu}\epsilon\phi
-\bar\psi\gamma^\mu\epsilon\,\sigma_1\phi +i\bar\psi\gamma^\mu\gamma^3\epsilon\,\sigma_2\phi
-i\frac{q}{2\ell}\bar\psi\gamma^\mu\epsilon\,\phi
\nn\\
&\qquad\qquad
+D^\mu\bar\phi\,\bar\epsilon\psi+D_\nu\bar\phi\,\bar\epsilon\gamma^{\mu\nu}\psi
-i\bar\psi\gamma^\mu\bar\epsilon\, F
\nn\\
&\qquad\qquad
-\bar\phi\lambda\gamma^{\mu}\bar\epsilon\phi
+\bar\phi\,\sigma_1\bar\epsilon\gamma^\mu\psi
-i\bar\phi\,\sigma_2\bar\epsilon\gamma^3\gamma^\mu\psi
+i\frac{q}{2\ell}\bar\phi\,\bar\epsilon\gamma^\mu\psi
\Bigr)\,.
\end{align}

Therefore, the supersymmetry variations of actions
can be written by surface terms, 
and actually they vanish if we assume that the Killing spinors satisfy
the conditions \eqref{2d_e_be} and the bulk fields satisfy the boundary conditions \eqref{2d-bc-a}-\eqref{2d_chiral_bc}.

%%%%%%%%%%%%%%%%%%%%%%%%%%%%%%%%%%%%%%%%%%%%%%%%%%%%%%%%%%%%%
\section{Cancellations in the 1-loop factor}
\label{can1}
In this Appendix, we will show how the 
cancellations in the 1-loop factor for the chiral multiplet
in the 3D SUSY gauge theories occur more precisely.
We can see that
\begin{eqnarray}
 0 &=& [ \Delta_\phi , h_\phi]= [ \Delta_\phi , v^\mu_\pm D_\mu ], \\
 0 &=& [ \Delta_\psi , h_\psi]= [ \Delta_\psi , v^\mu_\pm D_\mu ], \\
 && [ h_\phi, v^\mu_\pm D_\mu ]=\mp 2 v^\mu_\pm D_\mu, \
\end{eqnarray}
where $h_\phi/\ell \equiv i v^\mu D_\mu$ and 
$h_\psi /\ell \equiv i v^\mu D_\mu -\frac{1}{2 \ell} v^\mu \gamma_\mu$.
Thus, the eigenmodes $\Delta$ can be chosen as 
the eigenmodes of $h$.
We can easily see that 
$h_\psi \epsilon=- \epsilon$,
$h_\psi \bar\epsilon= \bar\epsilon$
and $h_\phi e^{i(m \varphi-n \chi)}= (m+n) e^{i(m \varphi-n \chi)}$.
Thus, for $\psi$ with $h_\psi \psi= h(\psi) \psi$,
we have $h_\phi \phi_1= (h(\psi) +1) \phi_1$ 
where $\phi_1 = \bar\epsilon \psi$.
On the other hand, for $\phi$ with $h_\phi \phi= h(\phi) \phi$,
we have $h_\psi \psi_\pm= (h(\phi) -1) \psi_\pm$
where 
$\psi_\pm \equiv \Bigl(\nu_\pm-\omega+\frac{1}{\ell}\Bigr)\,\epsilon\phi
-i\gamma^\mu \epsilon\, \da_\mu\phi$
where $\nu_\pm\equiv\omega\pm\sqrt{M^2+\omega^2}$\, with 
$\Delta_\phi \phi=M^2 \phi$.

If we construct $\psi_+$ from 
a $\phi_1$ corresponding to an eigenmode $\psi$ with the eigenvalue $\nu$, 
we can show that
\begin{eqnarray}
 \psi_+ =(h(\psi)/\ell +w-\nu ) \psi,
\end{eqnarray}
which is proportional to the original $\psi$
except $h(\psi)/\ell +w-\nu =0$.
We can see that 
the eigenmode with $h(\psi)/\ell =-w+\nu$ 
are the ``lowest'' modes for $h$ with fixed $\Delta$.

Conversely, if we construct $\phi_1$ from 
a $\psi_\pm$ corresponding to an eigenmode $\phi$ with the eigenvalue 
$M^2=\nu(\nu-2w)$, 
we can show that
\begin{eqnarray}
 \phi_1 =(\pm (\nu-w)+1/ \ell -h(\phi)/\ell ) \phi,
\end{eqnarray}
which is proportional to the original $\phi$
except $\pm (\nu-w)+1/ \ell -h(\phi)/\ell =0$.
Therefore, the unpaired modes are the lowest and 
highest modes for $h$ of $\psi$ and $\phi$, respectively.
(For $\bar\phi$ and $\bar\psi$, we can also show the same conclusion.)
This conclusion is, of course, consistent with the discussions in section \ref{3D-th}.

%%%%%%%%%%%%%%%%%%%%%%%%%%%%%%%%%%%%%%%%%%%%%%%%%%%%%%%%%%%%%

%%%%%%%%%%%%%%%%%%%%%%%%%%%%%%%%%%%%%%%%%%%%%%%%%%%%%%%%%%%%%

%%%%%%%%%%%%%%%%%%%%%%%%%%%%%%%%%%%%%%%%%%%%%%%%%%%%%%%%%%%%%
\end{document}